 \documentclass[]{aa}
 \usepackage{natbib}
 \usepackage{aas_macros}
 \usepackage{rotating}
 \bibpunct{(}{)}{;}{a}{}{,}
 \usepackage{graphicx}
 
\begin{document}

\title{Constraining the mass transfer in massive binaries through progenitor
evolution models of Wolf-Rayet+O binaries}

\author{Jelena Petrovic\inst{1,2} 
       \and
        Norbert Langer\inst{1}
	\and
	Karel A. van der Hucht\inst{3,4}
        }

\authorrunning{Petrovic et al.}
\titlerunning{WR+O progenitors}

\institute{Sterrenkundig Instituut,
           Universiteit Utrecht,
	   Princetonplein 5, NL--3584~CC Utrecht,
           The~Netherlands
	   \and
	   Astronomical Institute,
	   Radboud Universiteit Nijmegen, 
	   Toernooiveld 1, NL--6525~ED,
	   Nijmegen, The Netherlands
           \and
           SRON, Nationaal Instituut voor Ruimte Onderzoek, 
	   Sorbonnelaan 2, NL--3584~CA Utrecht, The Netherlands
	   \and
	   Sterrenkundig Instituut Anton Pannekoek, Universiteit van Amsterdam,
	   Kruislaan 403, NL--1098~SJ Amsterdam, The Netherlands
             }
\date{Received; accepted} %% leave as is
\offprints{J.Petrovic,\\ \email{petrovic@astro.uu.nl}}
%%%%%%%%%%%%%%%%%%%%%%%%%%%%%%%%%%%%%%%%%%%%%%%%%%%%%%%%%%%ABSTRACT
\abstract{
Since close WR+O binaries are the result of a strong interaction 
of both stars in 
massive close binary systems, they can be used to constrain
the highly uncertain mass and angular momentum budget
during the major mass transfer phase.
We explore the progenitor evolution of the three
best suited WR+O binaries HD\,90657, HD\,186943 and HD\,211853, 
which are characterized by
a WR/O mass ratio of $\sim$0.5 and periods of 6..10 days.
We are doing so at three different levels of approximation:
predicting the massive binary evolution through 
simple mass loss and angular momentum loss estimates, through
full binary evolution models with parametrized mass transfer 
efficiency, and through binary evolution models including rotation
of both components and a physical model which allows to
compute mass and angular momentum loss from the binary system
as function of time during the mass transfer process.
All three methods give consistently the same answers.
Our results show that, if these systems formed through stable mass transfer,
their initial periods were smaller than their current ones,
which implies that mass transfer has started during the 
core hydrogen burning phase of the initially more massive star.
Furthermore, the mass transfer in all three cases must have
been highly non-conservative, with on average only $\sim$10\%
of the transferred mass being retained by the mass receiving star.
This result gives support to our system mass and angular momentum
loss model, which predicts that, in the considered systems,
about 90\% of the overflowing matter is expelled by the  
rapid rotation of the mass receiver close to the $\Omega$-limit,
which is reached through the accretion of the remaining 10\%.
%We also investigate the influence of the initial mass ratio, 
%initial orbital period, WR mass
%loss rate and rotation on the initial WR mass in these systems. 
}
%%%%%%%%%%%%%%%%%%%%%%%%%%%%%%%%%%%%%%%%%%%%%%%%%%%%%%%%%%%%%%%%%% END
%TITLE
\maketitle
%%%%%%%%%%%%%%%%%%%%%%%%%%%%%%%%%%%%%%%%%%%%%%%%%%%%%%%%%%%%%%%introduction
\section{Introduction}\label{intro}
The evolution of a star in a binary system can differ significantly 
from that of an isolated one 
with the same mass and chemical composition. The physical processes that enter 
binary evolution are the gravitational and
radiation field from the companion, as well as the centrifugal force
arising from the rotation of the system.  But, most important, it is the 
evolution of the 
more massive component that will influence dramatically the evolution of 
the system. In certain evolutionary phases, mass transfer
from one star to another can occur, changing the fundamental properties
of both stars as well as their future evolution. 

The rotational properties of binary components may play a key role
in this respect.
The evolution of massive single stars can be strongly influenced by
rotation 
\citep{2000ApJ...544.1016H,2000A&A...361..101M}, 
and evolutionary models of rotating stars are
now available for many masses and metallicities.
While the treatment of the rotational processes in these models
is not yet in a final stage 
\citep[magnetic dynamo processes are just being included][]
{2004IAUS..215b,2003A&A...411..543M}, they provide
first ideas of what rotation can really do to a star.
Effects of rotation, as important they are in single stars,
can be much stronger in the components of close binary systems:
Estimates of the angular momentum gain of the accreting star
in mass transferring binaries show that critical rotation
may be reached quickly \citep{1981A&A...102...17P,2000A&A...362.1046L,2004A&A...419..623Y}.
In order to investigate this, we need binary evolution models
which include a detailed treatment of rotation in the stellar interior,
as in recent single star models. However, in binaries, tidal
processes as well as angular momentum and accretion need to be considered
at the same time. Some first such models are now available and
are discussed below.

Angular momentum accretion and the subsequent rapid rotation
of the mass gainer may be essential for some of the most exciting cosmic
phenomena, which may occur exclusively in binaries: Type~Ia supernovae,
the main producers of iron and cosmic yardsticks to measure the
accelerated expansion of the universe \citep{2004A&A...419..645Y,2004A&A...419..623Y},
and gamma-ray bursts from collapsars,
which the most recent stellar models with rotation and magnetic fields
preclude to occur in single stars 
\citep{gammapaper,2004IAUS..215b,2004IAUS..215d}.
For both, the Type~Ia supernova progenitors and the gamma-ray burst
progenitors, it is essential to understand how efficient the
mass transfer process is and on which physical properties it depends. 
Further exciting astrophysical objects
whose understanding is affected by our understanding of mass transfer comprise  
X-ray binaries \citep{1998A&A...330..201C} and Type~Ib and~Ic supernovae 
\citep{1992ApJ...391..246P}.

How much matter can stars accrete from a binary companion?
As mentioned above, non-magnetic accretion, i.e. accretion via a 
viscous disk or via ballistic impact, transports angular momentum and can lead to a strong
spin-up of the mass gaining star. For disk accretion, it appears plausible that
the specific angular momentum of the accreted matter corresponds to
Kepler-rotation at the stellar equator; this leads to a spin-up of
the whole star to critical rotation when its initial mass is increased
by about 20\% \citep{1981A&A...102...17P}. 
It appears possible that mass accretion continues in this situation,
as viscous processes may transport angular momentum 
outward through the star, the boundary layer, and the accretion
disk \citep{1991ApJ...370..597P}.
However, as the star is rotating very rapidly, its wind mass
loss may dramatically increase \citep{1997lbv..conf...83L,1998A&A...329..551L}, which may render
the mass transfer process inefficient.
 
Observations of massive post-mass transfer binary systems constrain this
effect. \citet{2003IAUS..212..275L} and \citet{2004IAUS..215c} points out that there is
evidence for both extremes occurring in massive close binaries, i.e. for
quasi-conservative evolution as well as for highly non-conservative evolution.
In the present study, we are interested in those binaries that contain a 
Wolf-Rayet and a main sequence O~star. 
We have chosen to focus on three WN+O systems (HD\,186943, HD\,90657 and HD\,211853) 
which have similar mass ratios ($\approx$0.5) and orbital periods ($6..10$ days). 
As clearly the two stars in these systems must have undergone
a strong interaction in the past, an understanding of their
progenitor evolution may be the key to constrain the mass transfer
efficiency in massive binaries: which fraction of the mass leaving
the primary star is accumulated by the secondary star during a mass
transfer event?

Evolutionary calculations of massive close binaries were performed by various authors.
General ideas about the formation of WR+O binary systems were given by
\citet{1967AcA....17..355P}, \citet{1967ZA.....66...58K}, \citet{1972NPhS..239...67V}.
\citet{1979A&A....73...19V} modelled the evolution of massive Case~B binaries
with different assumptions for mass and angular momentum loss from the binary system.
\citet{1982A&A...105..260V} computed evolutionary models of massive close Case~B binaries with primary
masses between $20 \,\mathrm{M}_{\odot}$ and $160 \,\mathrm{M}_{\odot}$. He concluded that most of the
WR primaries are remnants of stars initially larger than $40 \,\mathrm{M}_{\odot}$ and 
that the accretion efficiency in these systems should be very below 0.3
in order to fit the observations.
\citet{1992A&AS...94..453D} computed detailed models of massive Case~B binary
systems 
%ranging in initial masses from $9 \,\mathrm{M}_{\odot}$ to $40 \,\mathrm{M}_{\odot}$ and 
for initial mass ratios of 0.6 and 0.9, assuming 
an accretion efficiency of~0.5. 
\citet{1999A&A...350..148W} and \citet{2001A&A...369..939W} modelled 
massive binary systems mass range 12..60$\,\mathrm{M}_{\odot}$
assuming conservative evolution,
and \citet{wellsteinphd} presented the first rotating binary evolution 
models for initial masses of $\approx$15$\,\mathrm{M}_{\odot}$ and initial mass ratios $q$$\approx$1.

While it was realized through these models that different mass accretion
may be needed to explain different observations, these efforts did not
have the potential to explore the physical reasons for non-conservative
evolution. I.e., there is no reason to expect that the mass transfer efficiency
remains constant during the mass transfer process in a given binary system,
nor that its time-averaged value is constant for whole binary populations.

%Different values of $\beta$ are used in modelling by different authors.
%So far, most popular options in modelling of binary evolution are $\beta=1$ which
%corresponds to a conservative evolution
%(\cite{1969A&A.....1..167R}, \cite{1969A&A.....3...83K}, \cite{1971ARA&A...9..183P}, 
%\citet{wellsteinphd}) 
%and $\beta=0.5$ (\cite{1992A&AS...94..453D}, \cite{1992A&AS...96..653D}). 
%However, these are just assumptions and any value of $\beta$ 
%between $0$ and $1$ is {\it a prori} possible. It has been proposed by \citet{1991A&A...252..159V}
%that $\beta$ depends on mass ratio of the system.

It is not yet known which physical processes can expel matter from a binary system. 
\citet{1991A&A...252..159V} proposed that if a binary component is more 
massive than $\approx$40-50$\,\mathrm{M}_{\odot}$ it will go through an LBV phase of enhanced 
mass loss, which will prevent the occurrence of RLOF.
\citet{2003A&A...404..991D} investigated the possibility that radiation pressure from the 
secondary prevents the accretion. They found that even for moderate mass transfer rates
(5$\cdot$10$^{\rm -6} \,\mathrm{M}_{\odot}\rm~yr^{\rm -1}$) the wind and photon momenta 
which emerge from the accretion star can not alter the
dynamics of the accretion stream. Here, we follow the suggestion
that the effective mass accretion rate can be significantly decreased 
due to the spin-up of the mass receiving star \citep{wellsteinphd,
2003IAUS..212..275L,2004IAUS..215c,2004RMxAC..20..231P}.

%In our models, we investigate the possibility that the progenitor evolution of WR+O
%systems (HD\,186943, HD\,90657 and HD\,211853) was through Case~A mass transfer with
%$\beta\approx0.1$. For nonrotating models we assume $\beta=0.1$ based on the estimate
%from the simplified approach. For rotating models accretion efficiency decreases when the
%secondary star is spinning up close to critical rotation.
%Primary mass range for nonrotating models is $41..75 \,\mathrm{M}_{\odot}$ and for rotating models
%$41..65 \,\mathrm{M}_{\odot}$.

%Further more, there are other possibilities in binary evolution, which are so far, 
%not known and explained in details: common envelope  
%and ring structure. In the case of very 
%violent mass transfer, secondary can expand and fill it Roche lobe, and common
%envelope evolution will occur. Due to the friction of stars moving in envelope
%angular momentum will be decreased and system will end up in a merger or
%as a system with very small period \citep{1990ApJ...358..189D}.
%Also, system can go through a ring phase. After secondary fills its Roche lobe,
%in the case of not extremely high mass transfer rate, system will start losing
%mass through second Lagrangian point. Lost matter will form a ring around system
%and remove some of the angular momentum. System will end up similar as in the
%case of common envelope evolution \citep{1998A&ARv...9...63V}.

The remainder of this paper is organized as follows.
In Sect.~\ref{obs} we briefly discuss the observational data
available for WR+O binary systems. In Sect.~\ref{simple} 
we derive estimates for the masses of both stars in WR+O systems 
for given initial masses and accretion efficiencies.
In Sect.~\ref{code} we present the
physics used to compute our detailed evolutionary  models.
Non-rotating binary evolution models with an adopted constant mass accretion efficiency are presented in Sect.~\ref{nonrot}. Our 
rotating models in which the mass accretion efficiency is obtained
selfconsistently are discussed in Sect.~\ref{rot}. 
We briefly compare our models with observations in 
Sect.~\ref{comp}.
Conclusions are given in Sect.~\ref{concl}.
                         
\label{sec:introduction}
%%%%%%%%%%%%%%%%%%%%%%%%%%%%%%%%%%%%%%%%%%%%%%%%%%%%%%%%%%%%%%%%%%%%%%%%%%%
\section{Observational data}\label{obs}
There are about 20 observed Wolf-Rayet+O binary systems with known masses
of components in the catalogue of \citet{2001NewAR..45..135V}.
We have chosen to model three spectroscopic double-lined systems: 
HD\,186943 (WN3), HD\,90657 (WN5) and GP~Cep (WN6/WCE)have, since
they have similar mass ratios ($q$=$M_{\rm WR}/M_{\rm O}$$\approx$0.5) and orbital periods
(6..10 days). 

WN+O systems that also have short orbital periods are V444 Cyg, CX Cep,
CQ Cep, HD\,94546, HD\,320102 and HD\,311884.
V444 Cyg has period of 4.2 days
and can be result of stable mass transfer evolution, but since mass ratio
of this
system is $\sim$0.3 we did not include it in this paper.
Orbital periods of CX Cep and CQ Cep
are very short ($\sim$2 days) and these systems
are probably the result of a contact evolution.
HD\,94546 and HD\,320102 are systems with very low masses of WR and O components
(4$\,\mathrm{M}_{\odot}$+9$\,\mathrm{M}_{\odot}$ and 2.3$\,\mathrm{M}_{\odot}$+4.1$\,\mathrm{M}_{\odot}$ respectively)
and HD\,311884 is extremely massive WR+O binary system
(51$\,\mathrm{M}_{\odot}$+60$\,\mathrm{M}_{\odot}$). Recently,  an even more massive WR+O system has been observed 83$\,\mathrm{M}_{\odot}$+82$\,\mathrm{M}_{\odot}$
\citep{2004A&A...420L...9R,2004ApJ...611L..33B}.

\vspace{3mm}
\begin{table*}
\caption{\label{wro}Basic parameters of selected WN+O SB2 binaries}
\begin{tabular*}{\textwidth}{@{\extracolsep{\fill}}lccc}
\hline\hline
                            &                       &                      &                               \\
WR number                   & WR\,21                & WR\,127              & WR153 $^b$                    \\
%variable star name          & V398\,Car             & QY\, Vul             & GP\,Cep                       \\
HD number                   & HD\,90657             & HD\,186943           & HD\,211853                    \\
%$v$ (mag)                   & 9.67                  & 10.33                & 9.08                          \\
%$b$\,-\,$v$ (mag)           & 0.27                  & 0.15                 & 0.27                          \\
%$d$ (kpc)                   & 3.9                   & 4.4                  & 2.8                           \\
spectral type               & WN5+O4-6              & WN3+O9.5V            & WN6/WCE+O3-6I          \\
$p$ (days)                  & 8.2546\,$\pm$\,0.0001 & 9.5550               & 6.6887                \\
$e$                         & 0.04\,$\pm$\,0.03     & 0.07\,$\pm$\,0.04    & 0\,$+$\,0                     \\
$q$                         & 0.52                  & 0.47                 & 0.54                          \\
$a$\,sin\,$i$ (R$_\odot$)   & 37\,$\pm$\,3          & 39\,$\pm$\,6         & $>$\,35.2                     \\
$M$\,sin$^3$$i$ (M$_\odot$) & 8.4                   & 9.3                  &                               \\
$i$ ($^\circ$)              & 50\,$\pm$\,4          & 55\,$\pm$\,8         & 73                            \\
$M_{\rm WR}$ (M$_\odot$)    & 19                    & 17                   & $>$\,6                        \\
$M_{\rm O}$ (M$_\odot$)     & 37                    & 36                   & $>$\,21                              \\
%in cluster/ass.             &                       & Vul\,OB2:            & Cep\,OB1:                     \\
%in ring nebula              & {\sl IRAS shell}      & Anon(WR\,127), S\,92 & S\,132 (r)                    \\
                            &                       &                      &                               \\
\hline\hline
\end{tabular*}

\vspace*{3mm}
Notes: \\
$a$: all parameters from compilation of \citet{2001NewAR..45..135V}, unless noted
     otherwise. \\
$b$: \citet{2002ApJ...577..409D}. \\

\end{table*}

The mass ratio of a binary system is determined from its radial velocity
solution, with an error of 5-10\%.  However, to determine the exact value
of the masses of the binary components, the value of the inclination of the
system has to be known.  Without knowledge of the inclination, only minimum
masses of the components can be determined, i.e., $M$\,sin$^3$$i$.
\citet{1981ApJ...244..157M} determined the minimum mass for the WR star in
HD\,186943 to be 9-11$\,\mathrm{M}_{\odot}$.  \citet{1982ApJ...259..213N}
determined the masses of the components of HD\,90657 in the range
11-14$\,\mathrm{M}_{\odot}$ for the WN4 component and 21-28$\,\mathrm{M}_{\odot}$
for the O-type component.  The masses of the WR components in HD\,186943
and HD\,90657 given in Table~\ref{wro} have been determined by
\citet{1996AJ....112.2227L} on the basis of improved values for the
inclination of these systems. \citet{2002ApJ...577..409D} determined
minimum masses of the components of the system GP Cep.  Previously,
\citet{1996AJ....112.2227L} suggested values of $M_{\rm WR}$=15$\,\mathrm{M}_{\odot}$ and $M_{\rm O}$=27$\,\mathrm{M}_{\odot}$ for this system.

There is no obvious hydrogen contribution in the WR spectrum in any of
these systems \citep{1981ApJ...244..157M,1982ApJ...259..213N}.
\citet{1981ApJ...244..157M} showed that hydrogen absorption lines are
fairly broad in the spectrum of HD\,186943, equivalent to ${\rm v}$$sin$${\rm
i}$$\simeq$250$\rm~km~s^{-1}$, thus the O-type star is rotating much faster
than synchronously.

Beside the fact that the binary system GP Cep has a similar mass ratio and
period as the other two systems, it has some very different properties as
well.  The spectral type of the WR component in GP Cep is a combination of
WN and WC \citep[WN6/WCE][]{2002ApJ...577..409D}. Also,
\citet{1981ApJ...244..157M} showed that, next to the main period of
$\sim$6.69 days of the binary system GP Cep, radial velocities of
absorption lines vary also with a period of 3.4698 days.  He proposed that
GP Cep is a quadruple system, consisting of two pairs of stars, WR+O and O+O.
\citet{1990A&A...227..117P} suggested that in both pairs one component is a
WR star.  However, \citet{2002ApJ...577..409D} showed that there is only
one WR star in this quadruple system.

%%%%%%%%%%%%%%%%%%%%%%%%%%%%%%%%%%%%%%%%%%%%%%%%%%%%%%%%%%%%%%%%%%%%%%%%%%%%
\section{The simple approach}\label{simple}

If the initial binary system is very close (an initial period is of the order of few 
days), RLOF occurs while the primary is still in the core hydrogen burning phase and Case~A
mass transfer takes place (fast and slow phase). 
When the primary expands due to shell hydrogen burning, it
fills its Roche lobe and Case~AB mass transfer starts. During this mass transfer the primary 
star loses the major part of its hydrogen envelope. After Case~AB mass transfer, 
the primary is a helium core burning Wolf-Rayet star.
During all this time, the secondary is still a main sequence star, but with an increased mass
due to mass transfer.
When the initial binary period is of the order of one to few weeks, the primary fills its Roche lobe for the
first time during shell hydrogen burning and Case~B mass transfer takes place. The primary
loses most of its hydrogen envelope, becomes a WR star and the secondary is an O star with
an increased mass.
Case C mass transfer occurs when initial period is of the order of years.
The primary fills its Roche lobe during helium shell burning and mass transfer
takes place on the dynamical time scale.
This scenario is not likely for chosen systems, 
since some of the secondary stars in WR+O systems have been observed to rotate faster than
synchronously. This means that they have accreted some matter which increased their spin angular
momentum.

We constructed a simple method to quickly estimate the  
post-mass transfer parameters for a large number of binary systems
for a given accretion efficiency $\beta$.
This allows us to narrow the space of possible initial
parameters (primary mass, secondary mass and orbital period)
that allows the evolution into a specific observed WR+O systems. 
%For a given initial masses, period and accreting efficiency ($\beta$) 
%we estimate masses of the WR and O stars.

We considered binary systems with 
initial primary masses $M_{\rm 1,in}$=25..100$\,\mathrm{M}_{\odot}$ 
and secondaries masses $M_{\rm 2,in}$=25/1.7..100$\,\mathrm{M}_{\odot}$
with an initial period of $3$ days. We assumed that the primary is transferring matter
to the secondary until it reaches the mass of its initial helium core (Eq.~\ref{caseA}). 

Matter that is not accreted on the secondary leaves the system with 
the specific angular momentum which corresponds to the secondary's orbital 
angular momentum \citep{2001MNRAS.321..327K}, which is consistent with
our approach for mass loss from the binary system (cf. Sect.~\ref{code}). 
Stellar wind mass loss is neglected.

More massive initial primaries produce more massive WR stars (helium cores) in general,
but if the star is in a binary system that goes through mass transfer during hydrogen
core burning of the primary (Case~A), this depends also on other parameters:\\
-If the initial period is longer, 
mass transfer starts later in the primary evolution and the initial helium core of the 
primary is more massive.\\ 
-If the initial mass ratio is further from unity, 
the mass transfer rate from the primary reaches higher values and the initial helium core mass
is smaller.

We initially want to restrict ourselves to systems that undergo stable mass transfer, i.e. avoid
contact situations. \cite{2001A&A...369..939W} found that the limiting initial mass ratio for 
conservative Case~A binary system is $M_{\rm 1,in}/M_{\rm 2,in}$$\sim$1.55 and for conservative Case~B 
systems $\sim$1.25. Since we allow non-conservative evolution, 
we consider initial mass ratio $q$$\le$1.7 for Case~A and $q$$\le$1.4 for Case~B. 
The observed WR+O systems (HD\,186943, HD\,90657 and HD\,211853) all have very short orbital
periods, between 6 and 10 days. Since, the net effect of the Case~A+Case~AB, 
or Case~B is a widening of the orbit (if there is no contact), we have to assume that 
the initial periods need to be shorter than, or approximatively equal to the observed ones.
We adopted a minimum initial orbital period of
3 days to avoid that the primary fills its Roche lobe on the ZAMS.

We estimated the minimum initial helium core masses, which are obtained
by the earliest Case~A systems, for systems 
with a mass ratio of $M_{\rm 1,in}/M_{\rm 2,in}$=1.7 and an initial period of
3 days ($M_{\rm 1,in}$$\ga$41$\,\mathrm{M}_{\odot}$) from the detailed
evolutionary models shown later in this paper
(Sect.~\ref{nonrot}):
\begin{equation}\label{caseA}
M_{\rm WR,in}=0.24*M_{\rm 1,in}+0.27. 
\end{equation}
In this linear approximation, we neglected the influence of the initial mass ratio on 
the initial WR mass. It is shown in Sect.~\ref{subsecq} that this dependence becomes 
important only for initial mass ratios above $q$$\simeq$2.

For Case~B binaries, the initial WR mass does not depend on the initial period  
and the initial mass ratio of the system, since during core hydrogen burning, the primary evolves as a single
star, without any interaction with the secondary. 
We estimated the relation between initial main sequence mass and initial WR mass
as a linear fit from the Case~B binary systems with initial primaries 
$M_{\rm 1,in}$$\ga$18$\,\mathrm{M}_{\odot}$ \citep{1999A&A...350..148W}:
\begin{equation}\label{CaseB}
M_{\rm WR,in}=0.53*M_{\rm 1,in}-4.92.
\end{equation}

The minimum initial period for a system to evolve through Case~B mass transfer depends on the initial primary mass and 
the mass ratio. We can estimate, based on the radii of the primaries at the end
of the main sequence evolution, what would be the initial orbital separation necessary to avoid the primary filling its 
Roche lobe before shell hydrogen burning. 
From Kepler's law follows that the orbital separation is proportional to the mass ratio 
$a$$\sim$${q^{\rm -1/3}}$ and the initial primary mass $a$$\sim$${{M_{\rm 1}}^{\rm 1/3}}$.
Since $q$$\sim$1 we can neglect this dependence and estimate the initial period for which the
radius of the primary at the end of MS is equal to its Roche radius. 
For this estimate we do not take into account stellar wind that will widen the orbit and decrease
the masses. 
We conclude that the Case~B limiting initial orbital period
for 40$\,\mathrm{M}_{\odot}$ is $\sim$10 days, for 45$\,\mathrm{M}_{\odot}$$\sim$15 days and for
75$\,\mathrm{M}_{\odot}$$\sim$30 days. Since the result of stable Case~B mass transfer is widening 
of the orbit, it follows that (stable mass transfer) 
Case~B binary systems can not be progenitors of 
the observed systems HD\,186943, HD\,90657 and HD\,211853 whose orbital periods are shorter than
$10$ days. However, WR star masses resulting from Case~B evolution are practically the same as those from
very late Case~A evolution, which is still considered in our analysis.

We calculate binary systems for early Case~A ($p_{\rm in}$=3 days) and for late Case~A 
($p_{\rm in}$$\approx$${p_{\rm limit}}$). 

\begin{figure}
  \centering
   \includegraphics[width=\columnwidth]{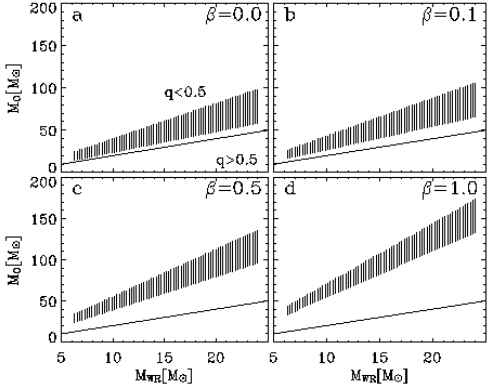}
  \caption{\label{casea}Masses of both components of post-Case~A mass transfer WR+O binary 
  systems resulting from our simple approach, for 
  initial primary masses in the range 25..100$\,\mathrm{M}_{\odot}$ and an initial 
  period of $p_{\rm in}$=3 days, for four different assumed accretion efficiencies
  $\beta$ (a-d). The solid line represents a mass ratio of
  $q$=$M_{\rm WR}/M_{\rm O}$=0.5. For an increasing $\beta$, the O stars in WR+O systems 
  become more massive and the WR/O-mass ratio decreases.}

\end{figure}

\begin{figure}
  \centering
   \includegraphics[width=\columnwidth]{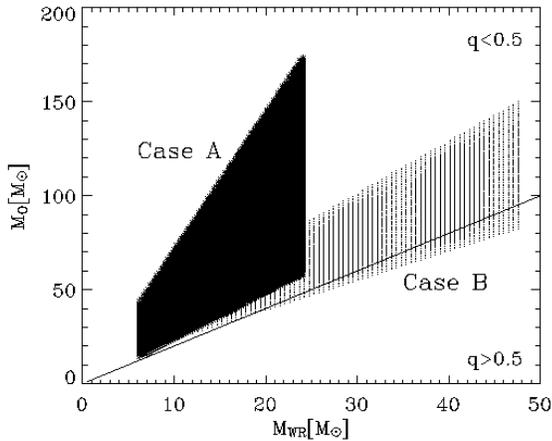}
  \caption{\label{caseab}
  Masses of both components of post-Case~A mass transfer WR+O binary
  systems resulting from our simple approach, for
  initial primary masses in the range 25..100$\,\mathrm{M}_{\odot}$ and
  for early ($p_{\rm in}$=3 days) Case~A and late Case~A respective Case~B evolution.
  The assumed accretion efficiency is $\beta$=0..1. 
  The solid line represents a mass ratio of
  $q$=$M_{\rm WR}/M_{\rm O}$=0.5.
  %WR stars are more massive for Case~B
  %evolution, since the progenitor primary star does not lose mass in mass transfer during core
  %hydrogen burning.
 }
  \end{figure}

The results are shown in Fig.~\ref{casea} for four different
accretion efficiencies ($\beta$=0.0,0.1,0.5,1.0 respectively) for early Case~A evolution
($p$=3 days). Fig.~\ref{caseab} shows the results for early Case~A systems ($p$=3 days) for 
all values of $\beta$=0..1 and for Case~B/late Case~A, also for all $\beta$.

We notice from Fig.~\ref{casea}, that when the assumed $\beta$ is larger, 
the resulting WR+O systems lie further from the line defined by $q$=0.5. The reason is clear:
if the accretion efficiency is higher, the secondary will become 
more massive while the initial mass of the WR star stays the same.
Conservative evolution (Fig.~\ref{casea}d) produces WR+O systems that have 
small mass ratios,
$q$=1/5..1/6.

We conclude that if the considered three observed WR+O 
binary systems evolved through a
stable mass transfer,
a large amount of matter must have left the system. On the other hand, 
since some of the secondary
stars in WR+O binaries have been observed to rotate faster than synchronously
\citep{1981ApJ...244..157M,1988PASP..100.1256U}, a certain amount of accretion
may be required.

Fig.~\ref{caseab} shows the resulting WR+O masses in Case~A and Case~B (latest Case~A) for accretion
efficiency $\beta$=0..1. If the primary star does not lose
mass in a mass transfer during core hydrogen burning ($p_{\rm in}$$\ge$$p_{\rm limit}$), 
it will form a more massive WR star, as we already explained.
There will be less mass to transfer from the primary to the secondary, and for fixed $\beta$
the corresponding O star will become less massive.
However, since the observed periods of HD\,186943, HD\,90657 and HD\,211853 are shorter than
10 days and Case~A+Case~AB widens the binary orbit, the initial orbital period should be shorter
than observed, so we can conclude roughly that $p_{\rm in}$ is between 3 and 10 days.

%This is also shown on
%Fig.~\ref{mwrq} for three examples:
%initial systems $41 \,\mathrm{M}_{\odot}+24\,\mathrm{M}_{\odot}$, $56 \,\mathrm{M}_{\odot}+33\,\mathrm{M}_{\odot}$ and  
%$75 \,\mathrm{M}_{\odot}+45\,\mathrm{M}_{\odot}$ ($\beta=0.1$). Upper plot shows initial WR and O mass for
%initial orbital periods $p_{\rm in}=3$ days and $p_{\rm in}=p_{\rm limit}$. Initial WR mass is 
%increasing and initial O mass is decreasing with increase of initial orbital period.
%Lower plot shows that mass ratio WR/O becomes larger if the initial binary system is wider.
%We do not know exact dependence of WR/O mass ratio on $P_{\rm in}$, but here we assumed that it is
%linear. We can conclude that mass ratio of $0.5$ can be reached for initial period larger than
%$3$ days. 

The orbital period of WR+O systems depends on their initial orbital period, 
their initial mass ratio and on the parameter $\beta$.
If the initial period increases and there is no contact during the evolution, 
the orbital period in the WR+O stage will also increase.
However, the orbital period of WR+O systems will be shorter if the initial mass ratio is larger.
If the initial masses are very similar, the primary will become less massive than the
secondary very early during the mass transfer, and afterward matter is
transfered from the less to the more massive star, which results in a widening of the orbit.
Conversely, the final period is shorter for a larger difference in initial masses in the
binary system. 

We can draw the following conclusions:\\
-The accretion efficiency during the major mass transfer phase 
in the progenitor evolution of the three observed WR+O binaries is small, i.e. 
$\beta$=0..0.1, as for larger $\beta$ the O~stars during the WR+O phase are more massive 
and the WR/O-mass rations smaller than observed.
However, we note that it is unlikely that the secondaries did not accrete at all ($\beta$=0),
since some O stars are found to rotate faster than synchronously. 
\\
-The initial orbital period needs to be larger than $\sim$3 days, 
to avoid contact at the beginning of hydrogen burning.\\
-The initial orbital period should be larger than $\sim$3 days, in order to
obtain massive enough WR stars.\\
-The initial orbital periods should be shorter than the observed orbital periods in the
three WR+O systems, i.e. shorter than $\sim$10 days. This excludes Case~B
mass transfer. 
\\
-While the initial mass ratio $M_{\rm 1,in}/M_{\rm 2,in}$ should not be too far from unity
so  contact is avoided, it should be close to the contact limit, since this leads
to the shortest orbital periods and largest WR/O mass ratios in WR+O systems, as needed 
for the three observed systems. 

%%%%%%%%%%%%%%%%%%%%%%%%%%%%%%%%%%%%%%%%%%%%%%%%%%%%%%%%%%%%%%%%%%%%%%%%%%%%%%%%%
\section{Numerical code and physical assumptions}\label{code}

We showed in Sect.~\ref{simple} that we can roughly estimate the parameters 
of the progenitor systems of observed WR+O binaries HD\,186943, HD\,90657 and HD\,211853. 
However, detailed numerical models are required in order to verify that 
the assumption of contact-free evolution can in fact be justified.
And finally, we want to check whether the required mass and angular momentum
loss can be reproduced by our detailed selfconsistent approach.

We are using a binary evolutionary code which was originally
developed by \citet{1998PhDT........29B} on the
basis of an implicit hydrodynamic stellar evolution code for single stars 
\citep{1991A&A...252..669L,1998A&A...329..551L}. 
It calculates simultaneous evolution of the two stellar
components of a binary system in a circular orbit
and the mass transfer within the Roche approximation
\citep{1978dcbs.conf.....K}.
Mass loss from the Roche lobe filling component through the first Lagrangian point
is given by \citet{1988A&A...202...93R} as:
\begin{equation}
{\dot M}={\dot M_{\rm 0}}\exp(R-R_{\rm l})/{H_{\rm p}}
\end{equation}
with ${\dot M_{\rm 0}}$=${{\rho}v_{\rm s}Q}/{\sqrt e}$,
where $H_{\rm p}$ is the photospheric pressure scale height, $\rho$ is the density, $v_{\rm s}$
the velocity of sound and $Q$ the effective cross-section of the stream through
the first Lagrangian point according to \citet{1983A&A...121...29M}.

Stellar wind mass loss for O stars on the main sequence is calculated according to
\citet{1989A&A...219..205K}. 
For hydrogen-poor stars ($X_{\rm s}$$<$0.4) we assume mass loss based on the empirical
mass loss rates for Wolf-Rayet stars derived by \citet{1995A&A...299..151H}:
\begin{equation}\label{maslos}
{\log({\dot M_{\rm WR}}/{\,\mathrm{M}_{\odot}\rm~yr^{\rm -1}})}=-11.95+1.5\log L/L_{\odot}-2.85X_{\rm s}.
\end{equation}
Since \citet{1998A&A...335.1003H} suggested that these mass loss rates may be
overestimated, we calculated evolutionary models with mass loss rate given by Eq.~\ref{maslos}
multiplied by factors 1/2, 1/3 and 1/6.

The treatment of a convection and a semiconvection which is applied here is 
described in \citet{1991A&A...252..669L} and \citet{1995A&A...297..483B}.  
Changes in chemical composition are computed using a nuclear network including pp chains,
the CNO-cycle, and the major helium, carbon, neon and oxygen burning reactions. 
More details are given in \citet{1999A&A...350..148W} and \citet{2001A&A...369..939W}.
We use the OPAL Rosseland-mean opacities of \citep{1996ApJ...464..943I}. 
For all models, a metallicity of $Z$=0.02 is adopted. The abundance ratios of
the isotopes for a given element are chosen to have the solar meteoritic abundance ratios
according to Grevesse \& Noels (1993).
The change of the orbital period (orbital angular momentum loss) due to the mass transfer and 
stellar wind mass loss is computed according to \citet{1992ApJ...391..246P}, 
with the specific angular momentum of the stellar wind material calculated by 
\citet{1993ApJ...410..719B}.
 
The influence of the centrifugal force in the rotating models 
is implemented according to \citet{1970stro.coll...20K}.
The stellar spin vectors are assumed to be perpendicular to the
orbital plane.  Synchronization due to tidal
spin-orbit coupling is included with a time scale given by \citet{1977A&A....57..383Z}.
Rotationally enhanced mass loss is included as follows:
\begin{equation}
{\dot M}/{\dot M (v_{\rm rot}=0)}={1}/{(1-{\Omega})^{\rm \xi}},
\end{equation}
where $\xi$=0.43, $\Omega$=$v_{\rm rot}/v_{\rm crit}$ and $v_{\rm crit}^{\rm 2}$=${GM(1-\Gamma)}/R$
with $\Gamma$=$L/L_{\rm Edd}$=${\kappa}L/(4{\pi}cGM)$ is Eddington factor, 
$G$ is gravitational constant, $M$
is mass, $R$ radius, $\kappa$ opacity, $v_{\rm rot}$ rotating velocity and 
$v_{\rm crit}$ critical rotational velocity \citep{1998A&A...329..551L}.

When the star approaches $\Omega$=1, the mass loss rate is increased according
to the previous equation. However, mass loss also causes a spin-down of the star
and  equilibrium mass loss rate $\Omega_{eq}$ results \citep{1998A&A...329..551L}.
If $\Omega>\Omega_{eq}$, the corresponding angular momentum loss is so large
that the star evolves away from the $\Omega$-limit.

The transport of angular momentum through the stellar interiour is formulated as a diffusive process:
\begin{eqnarray}
\left (\frac{\partial \omega}{\partial t} \right )_m={\frac{1}{i}} \left (\frac{\partial}{\partial
m} \right )_t \left [ \left (4\pi r^2 \rho \right )^2i \nu \left (\frac{\partial \omega}{\partial
m} \right )_t \right ] \nonumber\\
-{\frac{2w}{r}} \left (\frac{\partial r}{\partial t} \right )_m{\frac{1}{2}}{\frac{dlni}{dlnr}},
\end{eqnarray}
where $\nu$ is the turbulent viscosity and $i$ is the specific angular 
momentum of a shell at mass coordinate $m$.

The specific angular momentum of the accreted matter is determined by integrating the 
equation of motion of a test particle in the Roche potential in case the
accretion stream
impacts directly on the secondary star, and is assumed Keplerian otherwise \cite{wellsteinphd}.
Rotationally induced mixing processes and angular momentum transport through 
stellar interior are described by \citet{2000ApJ...528..368H}.
Magnetic fields generated due to differential rotation in the stellar interior
\citep{2002A&A...381..923S} are not included here \citep[however, see][]{gammapaper}.

%%%%%%%%%%%%%%%%%%%%%%%%%%%%%%%%%%%%%%%%%%%%%%%%%%%%%%%%%%%%%%%%%%%%%%%%%%%%%%%%%%
We calculated the evolution of the binary systems in detail until Case~AB
mass transfer starts. Then we estimated the outcome of this mass transfer 
by assuming that it ends when WR star has $\sim$5\% of the hydrogen left at the surface.
For this purpose we calculate the Kelvin-Helmholtz time scale of the primary:
\begin{equation}
{t_{\rm KH}=2\cdot10^{\rm 7}{M_{\rm 1}}^{\rm 2}/(L_{\rm 1}R_{\rm l1})\rm yr} 
\end{equation}
where $M_{\rm 1}$, $L_{\rm 1}$ and $R_{\rm l1}$ are mass, luminosity and Roche radius 
(in Solar units) of the primary star at the onset of Case~AB mass transfer.
The mass transfer rate is then assumed as:
\begin{equation}\label{mtr}
{\dot M_{\rm tr}=(M_{\rm 1}-M_{\rm WR,in})/{t_{\rm KH}}}
\end{equation}
where $M_{\rm WR,in}$ is the mass of the WR star that has a hydrogen surface abundance of 5\%; all quantities are
taken at the beginning of the mass transfer. 
We calculate the change of the orbital period orbit using
constant value of $\beta$=0.1 for non-rotating and $\beta$=0.0 for rotating models
\citep{wellsteinphd}.
Matter that is not retained by the secondary is assumed to leave the system with 
a specific angular momentum which corresponds to the secondary's orbital 
angular momentum \citep{2001MNRAS.321..327K}. 

%%%%%%%%%%%%%%%%%%%%%%%%%%%%%%%%%%%%%%%%%%%%%%%%%%%%%%%%%%%%%%%%%%%%%%%
\section{Non-rotating models}\label{nonrot}

We concluded in Sect.~\ref{simple} that massive O+O binaries can result in 
WR+O systems similar to observed the (HD\,186943,
HD\,90657 and HD\,211853) if accretion efficiency $\beta$ is low.
Since some O stars in WR+O binaries have been observed to rotate faster than synchronously, 
we concluded that $\beta$$>$0.0 and assumed a constant value of 
$\beta$=0.1 in our detailed evolutionary models. 
We already mentioned that the orbital periods of the observed systems are between 
6 and 10 days. Since the net effect of Case~A+Case~AB mass transfer is a widening 
of the orbit, the initial periods should be shorter than the observed ones, so 
we modelled binary systems with initial orbital periods of 3 and 6 days.

%%\begin{table}
\begin{sidewaystable*}
\caption[]{\label{big1}Non-rotating WR+O progenitor models for $\beta$=0.1.
N is the number of the model, $M_{\rm 1,in}$ and $M_{\rm 2,in}$ are initial masses of the primary and
the secondary, $p_{\rm in}$ is the initial orbital period and $q_{\rm in}$ is the initial mass ratio of
the binary system. $t_{\rm A}$ is time when Case~A mass transfer starts, $\Delta t_{\rm f}$
is the duration of the fast phase of Case~A mass transfer, $\dot M_{\rm tr}^{\rm max}$ is the maximum mass
transfer rate, $\Delta M_{\rm 1,f}$ and $\Delta M_{\rm 2,f}$ are mass loss of the primary and mass gain
of the secondary (respectively) during fast Case~A, $\Delta t_{\rm s}$ is the duration of slow Case~A mass
transfer,
$\Delta M_{\rm 1,s}$ and $\Delta M_{\rm 2,s}$ are mass loss of the primary and mass gain
of the secondary (respectively) during the slow Case~A, $p_{\rm AB}$ is the orbital period at the onset of Case
AB, $\Delta M_{\rm 1,AB}$ is the mass loss of the primary during Case~AB (mass gain of the secondary is
1/10 of this, see Sect.~\ref{code}), $M_{\rm WR,5}$ is the WR mass when the hydrogen surface abundance
is $X_{\rm s}$=0.05, the WR mass at $X_{\rm s}\le$0.01 is given in brackets, 
$M_{\rm O}$ is the mass of the corresponding O star, $q$ is the mass ratio $M_{\rm WR}/M_{\rm O}$, 
and
$p$ is the orbital period of the WR+O system.
The models are computed with a stellar wind mass loss of Hamann/6, except 
$^{\rm *}$ Hamann/3, $^{\rm **}$ Hamann/2.\\
$^{\rm c}$ indicates a contact phase that occurs
for low masses due to a mass ratio too far from unity, for high masses
due to the secondary expansion during slow phase of Case~A.}

\centerline{}
\begin{tabular}{lcccccccccccccccccc}
\hline
\hline
\\
$Nr$ & $M_{\rm 1,in}$ & $M_{\rm 2,in}$ & $p_{\rm in}$ & $q_{\rm in}$ & $t_{\rm A}$ & $\Delta t_{\rm f}$ & 
$\dot M_{\rm tr}^{\rm max}$ & $\Delta M_{\rm 1,f}$ & $\Delta M_{\rm 2,f}$ & $\Delta t_{\rm s}$ & 
$\Delta M_{\rm 1,s}$ & $\Delta M_{\rm 2,s}$ & $p_{\rm AB}$ &
$\Delta M_{\rm 1,AB}$ & $M_{\rm WR,5}(1)$ & $M_{\rm O}$ & $q$ & $p$\\
\\
\hline
$ $ & $\,\mathrm{M}_{\odot}$ & $\,\mathrm{M}_{\odot}$ & $\rm d$ & $$ & $10^{\rm 6} \rm yr$ & $10^{\rm 4} \rm yr$ & 
$\,\mathrm{M}_{\odot}/\rm yr$ & 
$\,\mathrm{M}_{\odot}$ & $\,\mathrm{M}_{\odot}$ & $10^{\rm 6} \rm yr$ & 
$\,\mathrm{M}_{\odot}$ & $\,\mathrm{M}_{\odot}$ & $\rm d$ &
$\,\mathrm{M}_{\odot}$ & $\,\mathrm{M}_{\odot}$ & $\,\mathrm{M}_{\odot}$ & $$ & $\rm d$\\
\hline
\\  
$N1$ & $41$ & $20$ & $3$ & $2.05$ & $2.8$ & $c$ & $-$ & $-$ & $-$ & 
$-$ & $-$ & $-$ & $-$ & $-$ & $-$ & $-$ & $-$ & $-$\\
\\
$N2$ & $41$ & $20$ & $6$ & $2.05$ & $3.6$ & $3.9$ & $5.4$ & $18.82$ & $1.85$ & 
$0.39$ & $0.97$ & $0.06$ & $5.9$ & $7.1$ & $11.8(11.2)$ & $22.5$ & $0.52$ & $12.6$ \\
\\
$N3$ & $41$ & $20.5$ & $3$ & $2.00$ & $2.8$ & $2.2$ & $18.0$ & $21.13$ & $2.11$ &
$-$ & $-$ & $-$ & $2.85$ & $8.17$ & $7.7(7.2)$ & $23.2$  & $0.33$ & $12.5$\\
\\
$N4$ & $41$ & $24$ & $3$ & $1.71$ & $2.8$ & $3.1$ & $3.6$ & $15.18$ & $1.51$ &
$1.51$ & $5.13$ & $0.20$ & $3.87$ & $9.05$ & $10.1(9.3)$ & $26.4$  & $0.38$ & $13.5$\\
\\
$N5$ & $41$ & $24$ & $6$ & $1.71$ & $3.6$ & $4.3$ & $3.2$ & $17.31$ & $1.72$ &
$0.42$ & $1.88$ & $0.09$ & $8.92$ & $7.53$ & $12.1(11.4)$ & $26.3$  & $0.46$ & $21.5$\\
\\
$N6$ & $41$ & $27$ & $3$ & $1.52$ & $2.75$ & $5.8$ & $1.9$ & $13.82$ & $1.37$ &
$1.51$ & $5.86$ & $0.17$ & $4.38$ & $9.59$ & $10.3(9.8)$ & $29.1$  & $0.35$ & $16.6$\\
\\
$N7$ & $41$ & $30$ & $3$ & $1.37$ & $2.7$ & $6.7$ & $1.1$ & $12.60$ & $1.24$ &
$1.51$ & $6.72$ & $0.08$ & $5.20$ & $9.76$ & $10.5(10.0)$ & $31.8$  & $0.33$ & $20.8$\\
\\
$N8$ & $45$ & $27$ & $3$ & $1.67$ & $2.5$ & $3.6$ & $3.3$ & $15.41$ & $1.53$ &
$1.57$ & $7.48$ & $0.25$ & $3.88$ & $8.81$ & $11.5(10.7)$ & $29.4$  & $0.39$ & $12.0$\\
\\
$N9$ & $56$ & $33$ & $3$ & $1.70$ & $1.9$ & $5.0$ & $4.1$ & $17.2$ & $1.70$ &
$1.86$ & $15.66$ & $0.44$ & $4.07$ & $7.14$ & $13.6(12.7)$ & $35.4$  & $0.38$ & $9.8$\\
\\
$N10$ & $56$ & $33$ & $6$ & $1.70$ & $2.8$ & $5.8$ & $3.1$ & $19.35$ & $1.9$ &
$0.60$ & $4.77$ & $0.02$ & $7.77$ & $9.18$ & $18.6(17.5)$ & $35.1$  & $0.53$ & $15.2$\\
\\
$N11^{\rm *}$ & $56$ & $33$ & $6$ & $1.70$ & $2.8$ & $5.8$ & $3.1$ & $19.35$ & $1.9$ &
$0.46$ & $3.63$ & $0.06$ & $7.91$ & $7.15$ & $18.6(17.2)$ & $34.9$  & $0.53$ & $13.8$\\
\\
$N12^{\rm **}$ & $56$ & $33$ & $6$ & $1.70$ & $2.8$ & $5.8$ & $3.1$ & $19.35$ & $1.9$ &
$0.43$ & $3.43$ & $0.07$ & $8.89$ & $3.5$ & $18.3(16.4)$ & $34.5$  & $0.53$ & $12.1$\\
\\
$N13$ & $65$ & $37$ & $3$ & $1.76$ & $1.6$ & $3.2$ & $4.7$ & $18.81$ & $1.87$ & 
$c$ & $-$ & $-$ & $-$ & $-$ & $16.2(14.8)$ & $-$ & $-$ & $-$\\
\\
$N14$ & $75$ & $45$ & $3$ & $1.67$ & $1.3$ & $4.2$ & $3.1$ & $18.57$ & $1.79$ & 
$c$ & $-$ & $-$ & $-$ & $-$ & $18.5(16.9)$ & $-$ & $-$ & $-$\\
\\
\hline
\hline
\end{tabular}
\normalsize
\end{sidewaystable*}
%%\end{table}

We chose initial primary masses to be in the range 41..75$\,\mathrm{M}_{\odot}$.
The masses of the secondaries are chosen so that the initial mass
ratio ($M_{\rm 1,in}/M_{\rm 2,in}$) is $q$$\approx$1.7-2.0. 
An initial mass ratio of $\approx$1.55 is estimated to be the limiting value 
for the occurrence of contact between the components 
in Case~A systems by \citet{2001A&A...369..939W} for conservative mass transfer.
Contact occurs when the accretion time scale of the secondary ($\dot M_{\rm 2,acc}/M_{\rm 2}$) 
is much longer than the thermal (Kelvin-Helmholtz) time scale of the primary 
($\dot M$=$M_{\rm 1}/t_{\rm KH}$), so the secondary expands 
and fills its Roche lobe. In our models, only $10$\% of matter lost by the primary 
is accreted on the secondary star, so it reaches hydrostatic equilibrium faster 
and expands
less than in the case of larger $\beta$.
This is the reason why we adopted a weaker condition for contact formation and    
calculate models with mass ratios $q$$\approx$1.7..2.0.

\begin{figure}
  \centering
   \includegraphics[width=\columnwidth]{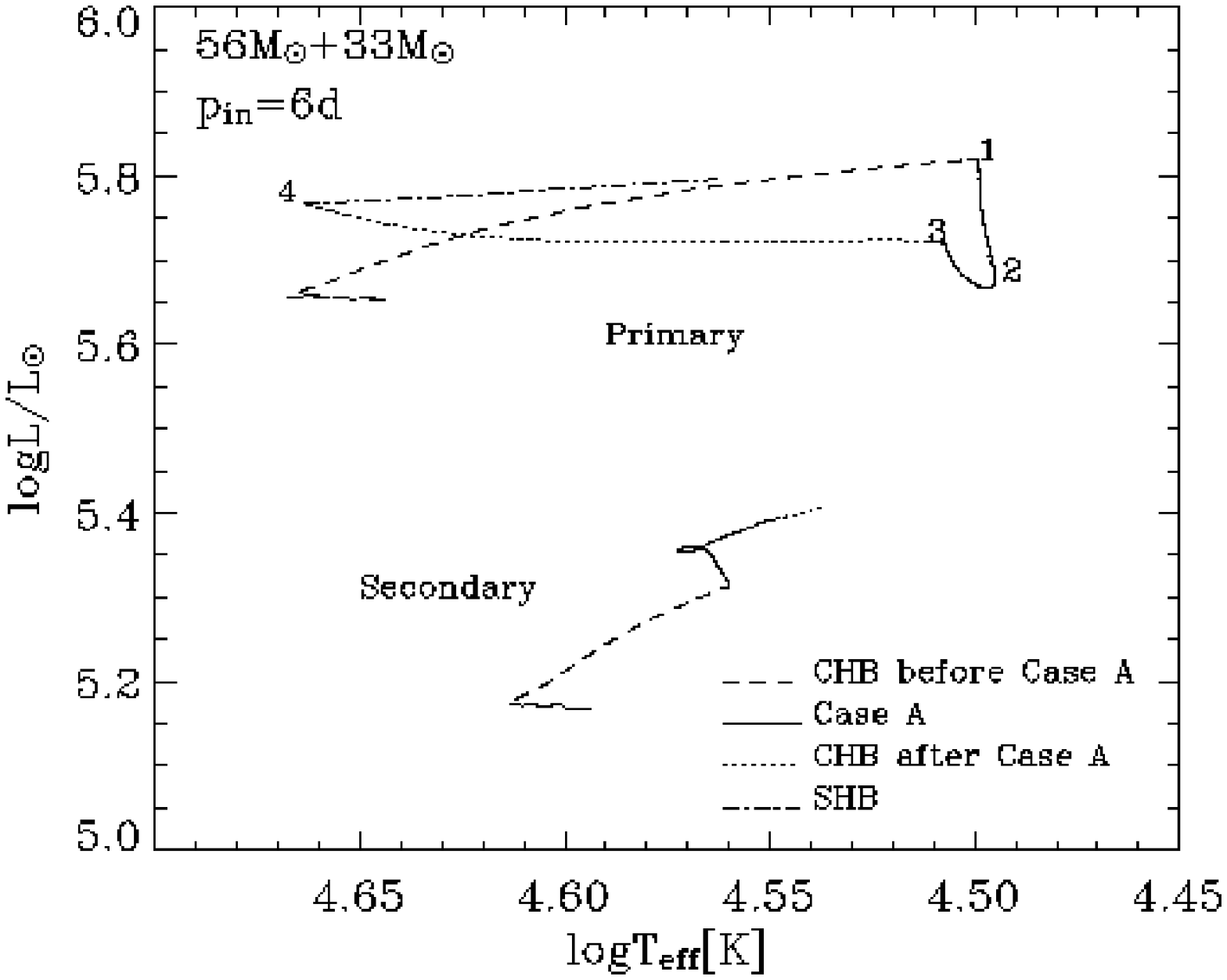}
  \caption{\label{hr}HR diagram of the initial system 
  $M_{\rm 1,in}$=56$\,\mathrm{M}_{\odot}$, $M_{\rm 2,in}$=33$\,\mathrm{M}_{\odot}$, $p_{\rm in}$=6 days.
  Both stars are core hydrogen burning (dashed line) until Case~A mass transfer starts (solid line).
  The primary is losing mass and its luminosity and effective temperature decrease. At the same
  time the secondary is accreting matter and expanding, becoming more luminous and cooler.
  After Case~A mass transfer is finished, the primary is losing mass by stellar wind and contracting 
  at the end of core hydrogen burning (dotted line). After this the primary starts with shell hydrogen
  burning and expands (dash-dotted line).
  }
\end{figure}

All modelled systems (except the ones that enter contact) go 
through Case~A and Case~AB mass transfer.
Details of the evolution of all calculated binary systems are given in Table~\ref{big1}.
We discuss the details of the binary evolution taking the system number 11 as an example.
Fig.~\ref{hr} shows the evolutionary tracks of the primary and the secondary in the HR
diagram until the onset of Case~AB mass transfer.
This system begins its evolution with the initial parameters 
$M_{\rm 1,in}$=56$\,\mathrm{M}_{\odot}$, $M_{\rm 2,in}$=33$\,\mathrm{M}_{\odot}$, $p_{\rm in}$=6 days.
Both stars are core hydrogen burning stars (dashed line, Fig.~\ref{hr}), but since the primary is more
massive, it evolves faster and fills its Roche lobe, so 
the system enters Case~A mass transfer (solid line, Fig.~\ref{hr}) 
$\sim$5.6$\cdot$10$^{\rm 6}$ years after the beginning of core hydrogen burning. 
The first phase of Case~A is fast process and takes place on the Kelvin-Helmholtz (thermal) time 
scale ($\sim$3.1$\cdot$10$^{\rm 4}$ years). The primary loses matter quickly and 
continuously with a high mass transfer rate
($\dot M_{\rm tr}^{\rm \rm max}$$\sim$3.1$\cdot$10$^{\rm
-3}\,\mathrm{M}_{\odot}\rm~yr^{\rm -1}$). 
In order to retain hydrostatic equilibrium, the envelope 
expands, which requires energy 
and causes a decrease in luminosity (Fig.~\ref{hr}). At the same time the secondary is 
accreting matter and is expanding.
Due to this, its luminosity increases and the effective temperature decreases (Fig.~\ref{hr}). 
During fast phase of Case~A mass transfer the primary loses $\sim$19$\,\mathrm{M}_{\odot}$ and the secondary
accretes 1/10 of that matter. 
After the fast process of mass transfer, 
the primary is still burning hydrogen in
its core and is still expanding, so  
slow phase of Case~A mass transfer takes place on a nuclear time scale (0.46$\cdot$10$^{\rm 6}$ years)
with a mass transfer rate of $\dot M_{\rm tr}$$\sim$10$^{\rm -6}
\,\mathrm{M}_{\odot}\rm~yr^{\rm -1}$. 
After this, the primary is the less massive star, 
with decreased hydrogen surface abundance.
Stellar wind mass loss of the primary increases when its surface becomes hydrogen poor
($X_{\rm s}$$<$0.4).
At the end of core hydrogen burning the primary contracts (effective temperature increases) 
and thus RLOF stops (Fig.~\ref{hr} dotted line). 
When the primary starts 
shell hydrogen burning it expands (dash-dotted line, Fig.~\ref{hr}), 
fills its Roche lobe and Case~AB mass transfer starts.

\begin{figure}
  \centering
   \includegraphics[width=\columnwidth]{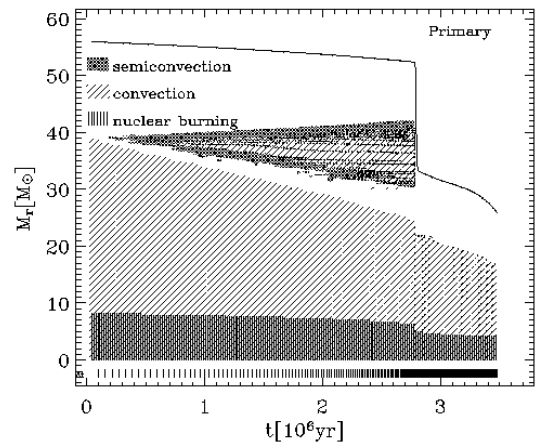}
  \caption{\label{conv1}The evolution of the internal structure of the 56$\,\mathrm{M}_{\odot}$ primary during the 
  core hydrogen
  burning. Convection is indicated with diagonal hatching and semiconvection with crossed hatching.
  The hatched area at the bottom indicates nuclear burning. 
  The topmost solid line corresponds to the surface of the star.}
  \label{fig5}
\end{figure}

\begin{figure}
  \centering
   \includegraphics[width=\columnwidth]{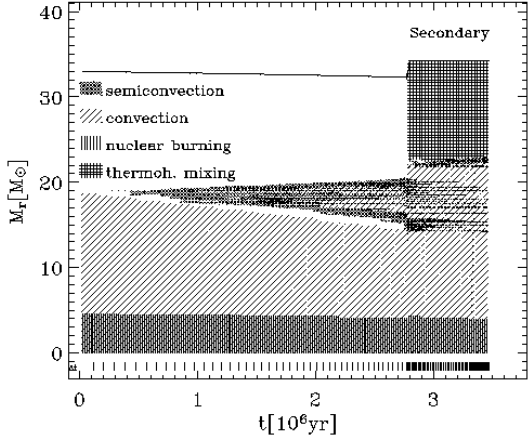}
  \caption{\label{conv2}The evolution of the internal structure of the 33$\,\mathrm{M}_{\odot}$ secondary during core hydrogen
  burning of the primary. Convection is indicated with diagonal hatching, semiconvection with
  crossed hatching and thermohaline mixing with 
  straight crossed hatching. 
  The hatched area at the bottom indicates nuclear burning. 
  The topmost solid line corresponds to the surface of the star.}
\end{figure}

\begin{figure}
  \centering
   \includegraphics[width=\columnwidth]{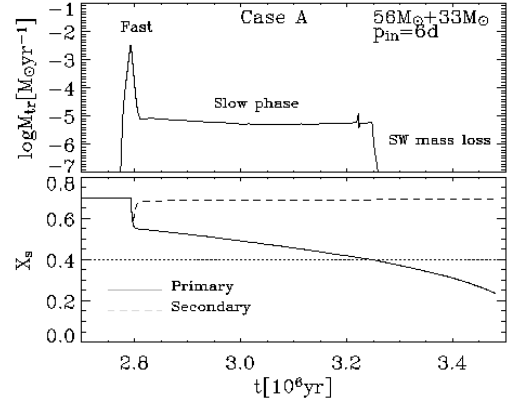}
  \caption{\label{surf}Upper plot: Mass transfer rate during Case~A mass transfer in the binary system
  with $M_{\rm 1,in}$=56$\,\mathrm{M}_{\odot}$, $M_{\rm
  2,in}$=33$\,\mathrm{M}_{\odot}$, $p_{\rm in}$=6 days.
  Lower plot: The hydrogen surface abundance of the primary (solid line) is decreasing during mass transfer
  and further due to stellar wind mass loss. The secondary (dashed
  line) recovered its original surface hydrogen abundance through thermohaline mixing. The primary
  starts losing mass with WR stellar wind mass loss when its hydrogen surface abundance falls beneath
  $X_{\rm s}$=0.4, represented by the dotted line. }
\end{figure}

\begin{figure}
  \centering
   \includegraphics[width=\columnwidth]{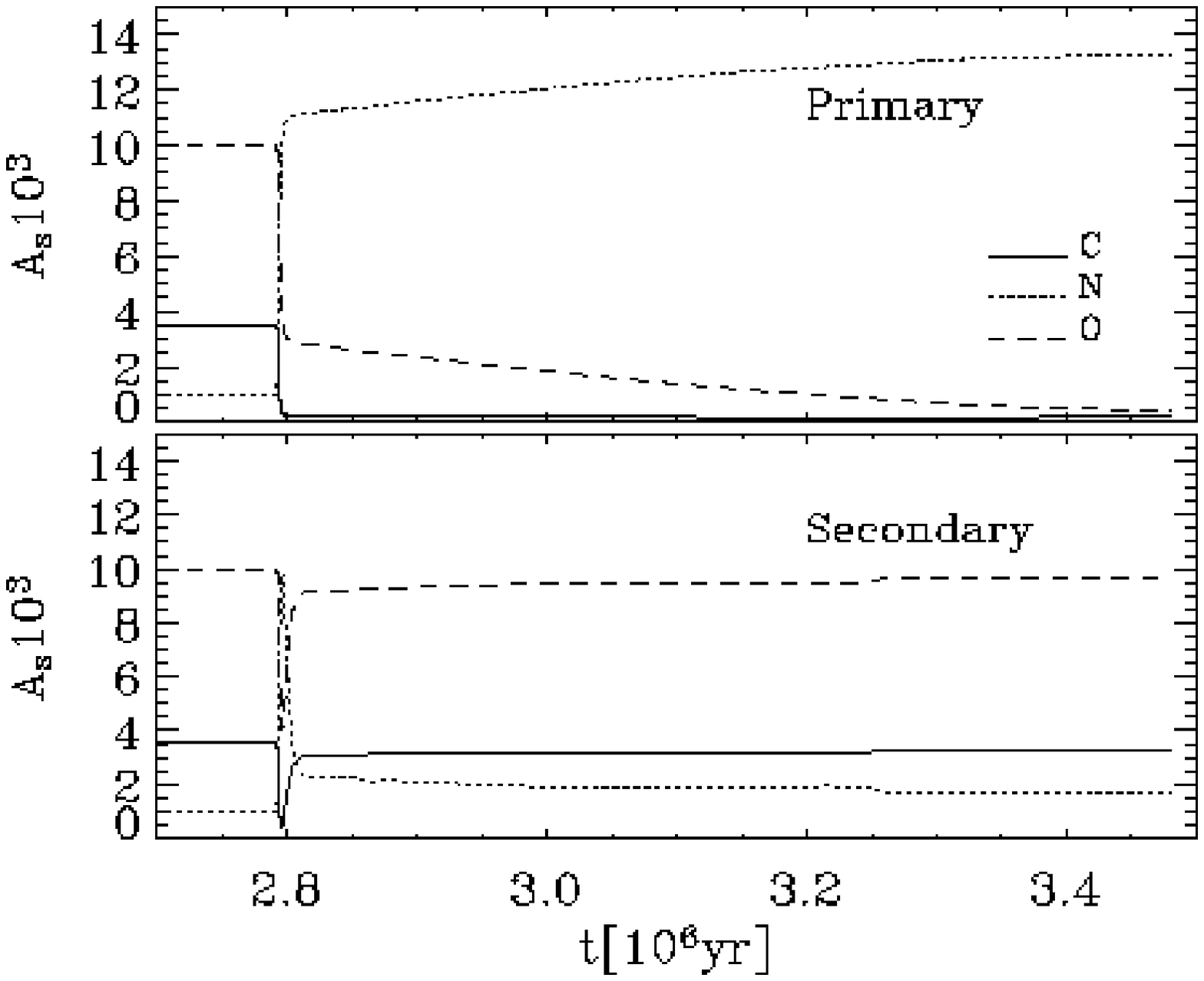}
  \caption{\label{cno}Surface abundance of carbon (solid line), nitrogen (dotted line) and 
  oxygen (dashed line) of the primary (upper plot) 
  and the secondary (lower plot) in the system with
  $M_{\rm 1,in}$=56$\,\mathrm{M}_{\odot}$, $M_{\rm 2,in}$=33$\,\mathrm{M}_{\odot}$, $p_{\rm in}$=6 days.}
\end{figure}

Fig.~\ref{conv1} and Fig.~\ref{conv2} show the evolution of the interior of the primary and the secondary until 
Case~AB mass transfer.
The primary loses huge amounts of matter during fast Case~A mass transfer and its convective core becomes less than a half of its original mass. At the same time, the
secondary accretes matter from the primary and the heavier elements are being relocated by thermohaline
mixing.
In Fig.~\ref{surf} and Fig.~\ref{cno} we see the mass transfer rate and 
the surface abundances of hydrogen, carbon, nitrogen and oxygen.

During Case~AB mass transfer the primary 
star loses the major part of its hydrogen envelope. After Case~AB mass transfer, 
the primary is a helium core burning star (WR) and the secondary is still 
a core hydrogen burning O star.
The masses of the modelled WR stars are in the range from $\sim$8..18.5$\,\mathrm{M}_{\odot}$ 
The orbital periods of the modelled WR+O systems vary 
from $\sim$9.5 to $\sim$20 days, and the mass ratios are between 0.33 and 0.53.

\subsection{Relation between initial and WR mass}\label{subsecmass}
The initial mass of helium core of the primary in the binary system depends on a
few parameters: initial primary mass, initial period, initial mass ratio and
stellar wind mass loss rate. 
If the primary loses matter due to the mass transfer or stellar wind during core hydrogen
burning,
it will form a helium core that is less massive than if
there was no mass loss. 
If the initial period is very short, Case~A mass transfer will take place
very early in the evolution of the primary, so the star will not have time to
develop a larger core before it starts losing mass due to Roche lobe overflow.
If the initial mass ratio ($q$=$M_{\rm 1,in}/M_{\rm 2,in}$) increases,
the mass transfer
rate from the primary star increases too and  
this results in less massive primaries that will
evolve into less massive WR stars.

\begin{figure}
  \centering
   \includegraphics[width=\columnwidth]{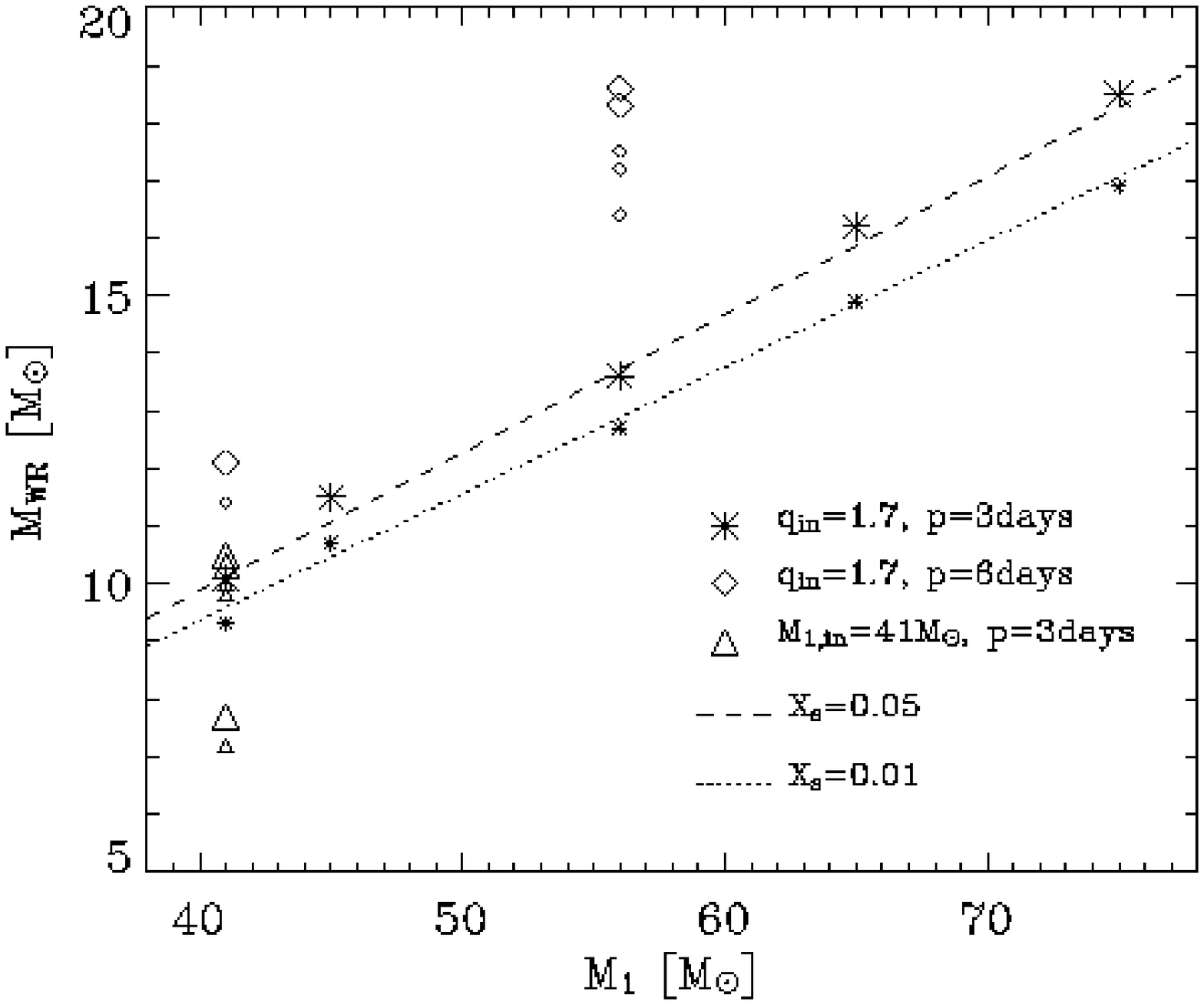}
  \caption{\label{mwr}Initial WR mass as a function of initial (progenitor) mass. 
  Large and small symbols indicate WR stars with hydrogen surface abundance of $X_{\rm s}$=0.05 and 
  $X_{\rm s}\le$0.01, respectively.
  Systems with an initial orbital period of $p_{\rm in}$=3 days and a mass ratio of $q\sim$1.7 
  are indicated with star symbols, systems with an initial period of $6$ days and $q\sim$1.7 with diamond symbols,
  systems with an initial primary 41$\,\mathrm{M}_{\odot}$ and initial period of $3$ days with triangle symbols.
  The dashed line represents a linear fit for systems with an initial period of $3$ days and 
  with $X_{\rm s}$=0.05, and the dotted line
  represents linear fit for systems with an initial period of 3 days and $X_{\rm s}\le$0.01.  }
\end{figure}

In our models the primary starts losing mass by stellar wind as WR star when its hydrogen surface
abundance goes below $X_{\rm s}$=0.4. However, the observed WR stars in HD\,186943, HD\,90657 and HD\,211853 do not have
obvious hydrogen on the surface, so we
assume that these WR stars are the result of Case~AB mass transfer, with
a hydrogen surface abundance of $X_{\rm s}$$\approx$0.05. We also calculated the 
corresponding WR masses with
$X_{\rm s}$$\le$0.01.
We plotted in Fig.~\ref{mwr} the initial WR masses ($X_{\rm s}$=0.05 and $X_{\rm s}\le$0.01) 
versus the initial primary (progenitor)
masses. With 'star' symbols we indicated WR stars that originate from binary
systems with an initial mass ratio of $q$$\approx$1.7 and an initial period
$p=3$ days (Table~\ref{big1}: N 4, 8, 9, 13, 14). Large 'star' symbols represent 
WR stars with 5\% of hydrogen at the
surface and small symbols indicate WR stars that
have a hydrogen surface abundance of less than 1\%. 

We derive a relation between the initial primary mass and the
initial WR mass (derived as a linear fit)
for $p$=3 days and $q$$\approx$1.7, ($X_{\rm s}$=0.05):
\begin{equation}
M_{\rm WR}=0.24*M_{\rm 1,in}+0.27.
\end{equation}
We use this relation to estimate the initial parameters of the possible
progenitors of the observed WR+O binary systems, as already explained in Sect.~\ref{simple}.
In the same way, the relation between the initial primary mass and the
initial WR mass ($X_{\rm s}$$<$0.01) for the same systems is:
\begin{equation}
M_{\rm WR}=0.22*M_{\rm 1,in}+0.56.
\end{equation}

We also show in Fig.~\ref{mwr} the initial WR masses ($X_{\rm s}$=0.05 for binary systems  
N 5, 10, 11, 12, Table~\ref{big1})
for an initial mass ratio of $\sim$1.7 and an initial orbital period
of $6$ days (diamond symbols). We notice that the resulting WR masses are higher than 
the ones that come out from systems with an initial
orbital period of 3 days (see Sect.~\ref{subsecper}). 
Different 'diamond' symbols for the initial primary 56$\,\mathrm{M}_{\odot}$ are for different mass
loss rates (see Sect.~\ref{subsecsw}).
Triangle symbols in Fig.~\ref{mwr} show the initial WR masses for constant initial primary mass,
$M_{\rm 1,in}$=41$\,\mathrm{M}_{\odot}$, but for different initial mass ratios (see Sect.~\ref{subsecq}) 

Note that the WR masses that are the result of early Case~A progenitor evolution 
are significantly lower than ones that are the result
of Case~B evolution \citep{1999A&A...350..148W}, because of the mass transfer
from the primary during the core hydrogen burning phase. 

\subsection{Influence of the initial mass ratio on the WR mass and orbital period}\label{subsecq}
\begin{figure}
  \centering
   \includegraphics[width=\columnwidth]{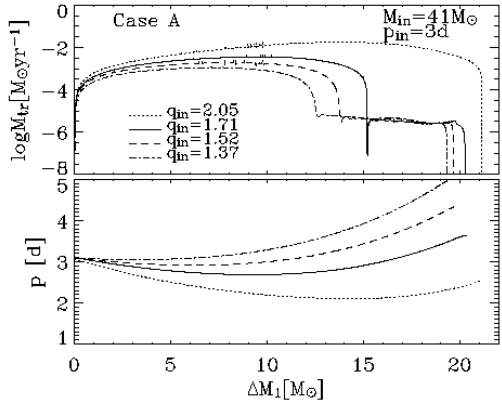}
  \caption{\label{masper}Mass transfer rate (upper plot) and orbital period (lower plot) during Case~A
  mass transfer as a function of the change of the primary mass for 
  systems with the initial primary $M_{\rm 1,in}$=41$\,\mathrm{M}_{\odot}$, initial orbital period
  $p_{\rm in}$=3 days and four different initial mass ratios: 2.00 (dotted line), 
  1.71 (solid line), 1.52 (dashed line) and 1.37 (dash-dotted line).
  (Table~\ref{big1} N 3, 4, 6, 7)}
\end{figure}

During the mass transfer phase, the mass transfer rate increases roughly until the 
masses of both components are equal. 
The maximum mass transfer rate during Case~A increases with the increase of the initial mass ratio 
($M_{\rm 1,in}/M_{\rm 2, in}$) and the resulting WR star is less massive.
To analyse the influence of the initial mass ratio on the evolution of the binary system, we compared
systems with an initial primary mass of $M_{\rm 1,in}$=41$\,\mathrm{M}_{\odot}$, an initial 
orbital period of
$p_{\rm in}$=3 days for five different initial mass ratios: 2.05, 2.00, 1.71, 1.52 and 1.37.
(Table~\ref{big1} N 1, 3, 4, 6, 7, Fig.~\ref{masper}).
The system with $q_{\rm in}$=2.05 enters contact during fast Case~A mass transfer. The 
mass transfer rate
in this case is very high ($\dot M$$\approx$6$\cdot$10$^{\rm -2}
\,\mathrm{M}_{\odot}\rm~yr^{\rm -1}$), 
the secondary expands, fills its Roche lobe and the system enters a contact phase.  
The system with an 
initial mass ratio of $q_{\rm in}$=2.00 loses $\sim$21$\rm
M_{\odot}$ during the fast phase of Case~A. The maximum mass transfer rate of this system is
$\dot M$$\approx$1.8$\cdot$10$^{\rm -2} \,\mathrm{M}_{\odot}\rm~yr^{\rm -1}$.
The helium surface abundance of the primary 
after this mass transfer is 65\%, so the 
primary shrinks, loses mass through a WR stellar wind 
and there is no slow phase of Case~A mass transfer ($R_{\rm 1}$$<9$$\,\mathrm{R}_{\odot}$).
For the other three models $q=$1.71,1.52,1.37, the primaries lose less mass 
($\sim$15,14,13$\,\mathrm{M}_{\odot}$ respectively) during the fast phase of Case
A mass transfer. The helium surface abundances in these systems after fast Case~A mass transfer 
are $\sim$30-35\%. The primaries
expand ($R$$\approx$12-15$\,\mathrm{R}_{\odot}$) on a nuclear time scale and transfer 
mass to the secondaries (slow phase of Case~A). 

We can conclude the following:
First, if the initial mass ratio is larger, the mass transfer rate from the primary during fast phase 
Case~A mass transfer is higher.
Second, if the mass transfer rate is higher, the helium surface abundance of the star
increases faster and if it reaches $\approx$58\%, the primary starts losing mass with
a higher (WR) mass loss rate and slow Case~A mass transfer can be avoided.

We also show in Fig.~\ref{masper} (lower plot) how the period changes during 
Case~A mass transfer for binary systems N 3, 4, 6, 7.
Roughly, when the mass is transfered from the more to the less massive star, the
binary orbit shrinks, and when the mass is transfered from the less to the more massive
star, the orbit widens.
If the initial period is close to unity, 
the absolute difference between stellar masses is small, and more mass is
transfered from the less to the more massive star during the evolution of the system.
This results in a longer final period after Case~A mass transfer.
Systems with initial mass ratios of 2.00, 1.71, 1.52 and 1.37 enter Case~AB mass
transfer with orbital periods of 2.9, 3.9, 4.4 and 5.2 days respectively.
However, the final period is also (more significantly) 
influenced by the stellar wind mass loss rate and the amount of matter lost
from the primary during Case~AB mass transfer (see
Section~\ref{subsecsw}).

\subsection{Influence of the initial period on the WR mass}\label{subsecper}
Depending on the initial orbital period of a binary system, Case~A mass transfer phase will
start earlier or later in the evolution. If the period is larger, the primary
will develop a larger core, before it starts transferring mass onto the secondary,
and the resulting WR star will be more massive. 
To investigate the influence of the initial period, we compare binary systems
41$\,\mathrm{M}_{\odot}$+24$\,\mathrm{M}_{\odot}$ and
56$\,\mathrm{M}_{\odot}$+33$\,\mathrm{M}_{\odot}$
with $p$=3 days and $p$=6 days (Table~\ref{big1}: N 4, 5, 9, 10).

\begin{figure}
  \centering
   \includegraphics[width=\columnwidth]{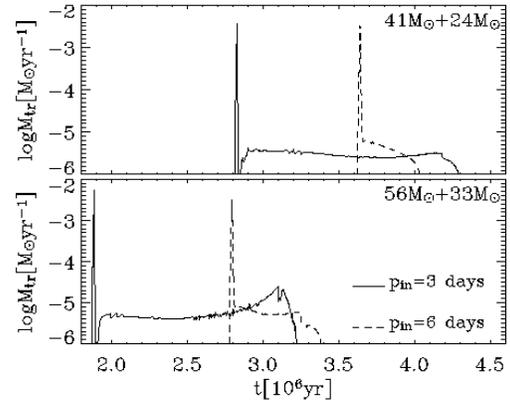}
  \caption{\label{p3p6}Mass transfer rate during Case~A mass transfer for systems
  41$\,\mathrm{M}_{\odot}$+24$\,\mathrm{M}_{\odot}$ (upper plot) and 56$\,\mathrm{M}_{\odot}$+33$\,\mathrm{M}_{\odot}$ (lower plot)
  and an initial orbital period of $p_{\rm in}$=3 days(solid line) and $p_{\rm in}$=6 days (dashed line).
  Case~A mass transfer starts later in initially wider binary systems, the primary has more time to
  increase mass of its core and the initial WR star is more massive (Table~\ref{big1}: N 4, 5, 9, 10).}
  \label{fig4}
\end{figure}
If the initial orbital period increases for 3 days, a 41$\,\mathrm{M}_{\odot}$ star will 
enter Case~A mass transfer $\sim$8$\cdot$10$^{\rm 5} \rm~yr$ later and a 56$\,\mathrm{M}_{\odot}$ 
star $\sim$9$\cdot$10$^{\rm 5} \rm~yr$ later. 
So, there are two things to point out:
first, the more massive star (56$\,\mathrm{M}_{\odot}$) evolves faster, and second,
a 3 days longer 
initial period postpones Case~A mass transfer, for this star, by about 10$^{\rm5} \rm~yr$ 
more than for a 41$\,\mathrm{M}_{\odot}$ star.
The net effect is a more significant increase of the convective 
core (i.e. initial helium core, i.e initial WR mass),
for more massive star, due to the initial orbit widening.

\subsection{Influence of WR mass loss rate on masses and final period}\label{subsecsw}
In our models, we assume that 
when the star has less than 40\% of hydrogen at the surface, 
it starts losing mass according to the \citet{1995A&A...299..151H} WR mass loss rate,
multiplied by factors: 1/6, 1/3 and 1/2 (Table~\ref{big1}: N 10, 11, 12).
If the primary is losing more mass by stellar wind during core hydrogen burning, 
it will develop a less massive helium core. At the same time there will be less matter 
to be transfered during Roche lobe overflow, so the secondary will accrete less. 

\begin{figure}
  \centering
   \includegraphics[width=\columnwidth]{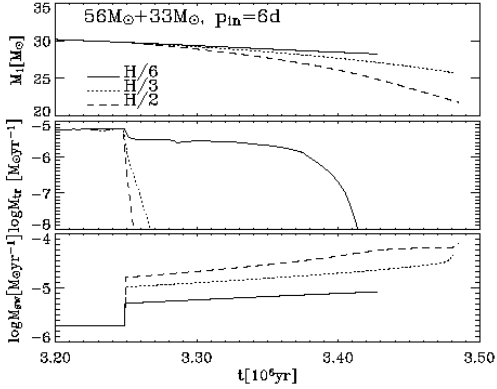}
  \caption{\label{mass}Primary mass (first plot), mass transfer rate (second plot) and stellar wind
  mass loss rate from the primary (third plot) until the onset of Case~AB mass transfer for the 
  system $M_{\rm 1,in}$=56$\,\mathrm{M}_{\odot}$, $M_{\rm 2,in}$=33$\,\mathrm{M}_{\odot}$, $p_{\rm in}$=6
  days and three different stellar wind mass loss rates (from $X_{\rm s}\le$0.4): 1/6 (solid line),
  1/3 (dotted line) and 1/2 (dashed line) of mass loss rate proposed by \citet{1995A&A...299..151H}
  (Table~\ref{big1}: N 10, 11, 12).}
  \label{fig4}
\end{figure}

We show in Fig.~\ref{mass} the influence of the stellar wind mass loss (Plot c) on the primary mass 
(Plot a) and mass transfer rate (Plot b) of
systems with $M_{\rm 1,in}$=56$\,\mathrm{M}_{\odot}$, $M_{\rm 2,in}$=33$\,\mathrm{M}_{\odot}$, $p_{\rm in}$=3
and three different stellar wind mass loss rates (from $X_{\rm s}$$\le$0.4): 1/6 (solid line),
1/3 (dotted line) and 1/2 (dashed line) of the mass loss proposed by \citet{1995A&A...299..151H}.  
We notice that for higher mass loss rates, the slow phase of Case~A
stops earlier, due to the decrease of the stellar radius. 
The orbit is widening due to the stellar wind mass loss and
the final period increases with the increasing mass loss rate.
However, the orbit is more significantly widening during Case~AB mass transfer. 
The more mass there is to
transfer from the primary to the secondary during Case~AB mass transfer, 
the larger the final orbital period.
So, if the stellar wind removes most of the hydrogen envelope of the primary,
there will be less mass to transfer during Case~AB and 
the net effect of a higher mass loss rate is a shorter orbital period of the WR+O system.

\section{Rotating models}\label{rot}

When mass transfer in a binary system starts, the primary loses matter through the first Lagrangian 
point ($L_{\rm 1}$). This matter carries a certain angular momentum that will be transfered to the
secondary. If there is an accretion disk, the angular momentum of the transfered matter is assumed to
be Keplerian. If there is a direct impact accretion, like in our models, we calculate the angular
momentum following a test particle moving through $L_{\rm 1}$.
This angular momentum spins up the top layers of the secondary star, and
angular momentum is transfered further into the star due to rotationally induced mixing processes.
Every time the secondary spins up to close to critical rotation it starts losing
more mass due to the influence of centrifugal force (Eq.~5). High mass loss decreases the net 
accretion efficiency and 
also removes angular momentum from the secondary star. The secondary star is also spun down by tidal
forces that tend to synchronize it with the orbital motion. \citet{wellsteinphd}
investigated these processes in binary systems with
initial mass ratios close to unity and concluded that
the accretion efficiency does not decrease significantly for Case~A mass transfer, but in
the Case~B the parameter $\beta$ can be significantly decreased by rotation. 
We present Case~A rotating models with larger mass ratio $q$=$M_{\rm
1,in}/M_{\rm 2,in}$=1.7..2 and 
find
that accretion can be significantly decreased during Case~A mass transfer. The reason is the following:
if the initial mass ratio increases, so does the maximum mass transfer rate ($\dot M_{\rm mtr}$
increases roughly until the masses in binary system are equal). If there is more mass  
transfered from the primary to the secondary, the rotational velocity of the secondary is higher
as well as its mass loss, which leads to a smaller accretion efficiency.

\begin{sidewaystable*}
\caption[]{\label{big2}Rotating WR+O progenitor models.
N is the number of the model, $M_{\rm 1,in}$ and $M_{\rm 2,in}$ are initial masses of the primary and
the secondary, $p_{\rm in}$ is the initial orbital period and $q_{\rm in}$ is the initial mass ratio of
the binary system. $t_{\rm A}$ is the time when Case~A mass transfer starts, $\Delta t_{\rm f}$
is the duration of the the fast phase of Case~A mass transfer, $\dot M_{\rm tr}^{\rm max}$ is the maximum mass
transfer rate, $\Delta M_{\rm 1,f}$ and $\Delta M_{\rm 2,f}$ are mass loss of the primary and mass gain
of the secondary (respectively) during the fast Case~A, $\Delta t_{\rm s}$ is the duration of slow Case~A mass
transfer,
$\Delta M_{\rm 1,s}$ and $\Delta M_{\rm 2,s}$ are mass loss of the primary and mass gain
of the secondary (respectively) during the slow Case~A, $p_{\rm AB}$ is the orbital period at the onset of Case
AB, $\Delta M_{\rm 1,AB}$ is the mass loss of the primary during Case~AB (mass gain of the secondary is
$1/10$ of this, see Sect.~\ref{code}), $M_{\rm WR,5}$ is the WR mass when the hydrogen surface abundance is
$X_{\rm s}=0.05$, $M_{\rm O}$ is the mass of the corresponding O star, $q$ is the 
mass ratio $M_{\rm WR}/M_{\rm O}$,
$p$ is the orbital period of the WR+O system and $M_{\rm WR,1}$ is WR mass with $X_{\rm s}/le0.01$.
The models are computed with a stellar wind mass loss of Hamann/6 except :$^{\rm *}$ Hamann/3, 
$^{\rm **}$ Hamann/2.\\
$^{\rm c}$ indicates a contact phase}

\centerline{}
\begin{tabular}{lccccccccccccccccccc}
\hline
\hline
\\

$Nr$ & $M_{\rm 1,in}$&$M_{\rm 2,in}$ & $p_{\rm in}$ & $q_{\rm in}$ & $t_{\rm A}$ & $\Delta t_{\rm f}$ & 
$\dot M_{\rm tr}^{\rm max}$ & $\Delta M_{\rm 1,f},\Delta M_{\rm 2,f}$ & $\Delta t_{\rm s}$ & 
$\Delta M_{\rm 1,s},\Delta M_{\rm 2,s}$ & $p_{\rm AB}$ &
$\Delta M_{\rm 1,AB}$ & $M_{\rm WR,5}(1),M_{\rm O}$ & $q$ & $p$\\
%$Nr$ & $M_{\rm 1,in},$ & $p_{\rm in}$ & $q_{\rm in}$ & $t_{\rm A}$ & $\Delta t_{\rm f}$ & 
%$\dot M_{\rm tr}^{\rm max}$ & $\Delta M_{\rm 1,f},$ & $\Delta t_{\rm s}$ & 
%$\Delta M_{\rm 1,s},$ & $p_{\rm AB}$ &
%$\Delta M_{\rm 1,AB}$ & $M_{\rm WR,5}(1),$ & $q$ & $p$\\
%& $M_{\rm 2,in}$ & & & & & & $\Delta M_{\rm 2,f}$ & & $\Delta M_{\rm 2,s}$ & & & $M_{\rm O}$  & & \\
\\
\hline

$ $ & $\,\mathrm{M}_{\odot}$ & $\,\mathrm{M}_{\odot}$ & $\rm d$ & $$ & $10^{\rm 6} \rm yr$ & $10^{\rm 4} \rm yr$ & 
$\,\mathrm{M}_{\odot}/ \rm yr$ & 
$\,\mathrm{M}_{\odot}$  & $10^{\rm 6} \rm yr$ & 
$\,\mathrm{M}_{\odot}$ & $\rm d$ & $\,\mathrm{M}_{\odot}$ &
$\,\mathrm{M}_{\odot}$  & $$ & $\rm d$\\
    
\hline
\\
$R1$ & $41$&$20$ & $6$ & $2.05$ & $3.4$ & $1.5$ & $6.5$ & $18.67,3.33$  &
$0.58$ & $2.38(1.37),0.81$ & $3.97$ & $6.61$ & $11.0(10.2),23.98$ & $0.46$ & $9.78$\\
\\
$R2^{\rm **}$ & $41$&$20$ & $6$ & $2.05$ & $3.4$ & $1.5$ & $6.5$ & $18.67,3.33$  & 
$0.10$ & $0.32(0.11),0.20$ & $4.77$ & $2.96$ & $10.4(9.0),23.20$ & $0.45$ & $7.92$\\
\\
$R3$ & $41$&$24$ & $3$ & $1.71$ & $2.6$ & $1.5$ & $3.9$ & $15.47,5.04$  & 
$1.34$ & $9.38(1.00),7.54$ & $4.27$ & $6.32$ & $8.2(7.6),36.17$ & $0.23$ & $17.86$\\
\\
$R4$ & $41$&$24$ & $6$ & $1.71$ & $3.4$ & $2.6$ & $3.8$ & $17.75,4.06$  &
$0.68$ & $2.55(0.9),1.53$ & $5.66$ & $7.25$ & $11.2(10.5),29.27$  & $0.38$ & $16.42$\\
\\
$R5$ & $56$&$33$ & $6$ & $1.70$ & $2.4$ & $3.7$ & $3.2$ & $19.32,2.91$  &
$0.98$ & $11.93(4.13),6.98$ & $6.09$ & $4.88$ & $14.9(13.6),42.09$  & $0.35$ & $11.59$\\
\\
$R6^{\rm *}$ & $56$&$33$ & $6$ & $1.70$ & $2.4$ & $3.7$ & $3.2$ & $19.32,2.91$ &
$0.90$ & $10.93(6.42),3.91$ & $6.64$ & $1.8$ & $14.8(12.8),38.99$  & $0.38$ & $8.53$\\
\\
$R7^{\rm **}$ & $56$&$33$ & $6$ & $1.70$ & $2.4$ & $3.7$ & $3.2$ & $19.32,2.91$ &
$0.45$ & $3.24(0.65),2.27$ & $8.43$ & $0.0$ & $11.2(8.8),37.04$  & $0.30$ & $8.43$\\
\\
$R8$ & $60$&$35$ & $6$ & $1.71$ & $2.3$ & $2.2$ & $3.0$ & $19.97,3.98$  &
$0.92$ & $12.32(4.20),7.21$  & $6.58$ & $5.27$ & $15.7(14.6),45.13$  & $0.35$ & $12.75$\\
\\
$R9^{\rm *}$ & $60$&$35$ & $6$ & $1.71$ & $2.3$ & $2.2$ & $3.0$ & $19.97,3.98$  &
$0.84$ & $11.42(6.26),4.58$ & $7.64$ & $0.0$ & $14.9(12.2),42.43$  & $0.35$ & $7.64$\\
\\
\hline
\hline
\end{tabular}
\normalsize
\end{sidewaystable*}

\begin{figure}
  \centering
   \includegraphics[width=\columnwidth]{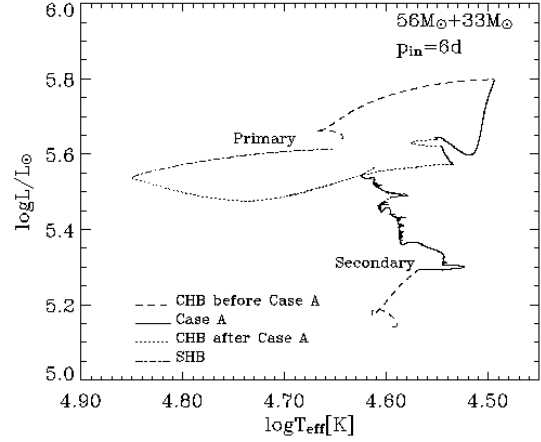}
  \caption{\label{hrrot}HR diagram of the initial system 
  $M_{\rm 1,in}$=56$\,\mathrm{M}_{\odot}$, $M_{\rm 2,in}$=33$\,\mathrm{M}_{\odot}$, $p_{\rm in}$=6 days with rotation.
  Both stars are core hydrogen burning (dashed line) until Case~A mass transfer starts (solid line).
  The primary is losing mass and its luminosity decreases. At the same
  time the secondary is accreting matter and expanding, becoming more luminous.
  After Case~A mass transfer is finished, the primary is losing mass by stellar wind and contracting 
  at the end of core hydrogen burning (dotted line). After this the primary starts shell hydrogen
  burning and expands (dash-dotted line).}
\end{figure}

\begin{figure}
  \centering
   \includegraphics[width=\columnwidth]{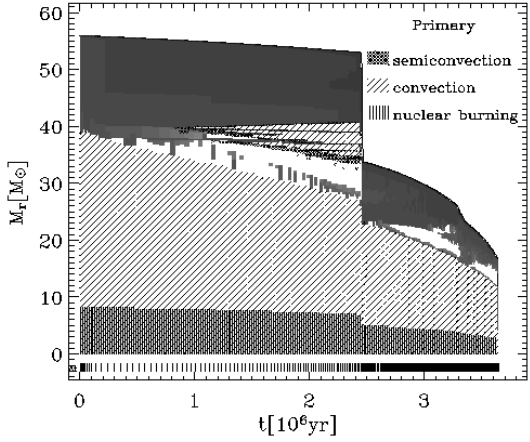}
  \caption{\label{conv1rot}The evolution of the internal structure of the rotating 56$\,\mathrm{M}_{\odot}$ 
  primary during core hydrogen
  burning. Convection is indicated with diagonal hatching and semiconvection with crossed hatching.
  The hatched area
  at the bottom indicates nuclear burning. 
  Gray shaded areas represent regions with rotationally induced
  mixing (intensity is indicated with different shades, the darker the colour, the stronger rotational
  mixing). The topmost solid line corresponds to the surface of the star.}
\end{figure}

\begin{figure}
  \centering
   \includegraphics[width=\columnwidth]{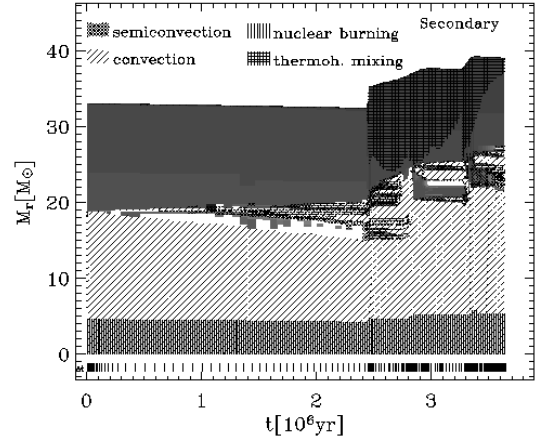}
  \caption{\label{conv2rot}The evolution of the internal structure of the rotating 33$\,\mathrm{M}_{\odot}$ 
  secondary during core hydrogen
  burning of the primary. Convection is indicated with diagonal hatching, semiconvection with
  crossed hatching and thermohaline mixing with straight crossed hatching. 
  The hatched area
  at the bottom indicates nuclear burning. 
  Gray shaded areas represent regions with rotationally induced
  mixing (intensity is indicated with different shades, the darker the colour, the stronger rotational
  mixing). The topmost solid line corresponds to the surface of the star.}
\end{figure}

\begin{figure}
  \centering
   \includegraphics[width=\columnwidth]{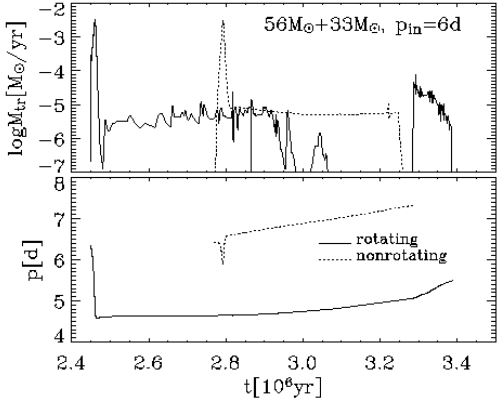}
  \caption{\label{mtrrot}Upper plot: The mass transfer rate during 
  Case~A mass transfer in the binary systems with
  $M_{\rm 1,in}$=56$\,\mathrm{M}_{\odot}$, $M_{\rm 2,in}$=33$\,\mathrm{M}_{\odot}$, $p_{\rm in}$=6 days with (solid line)
  and without rotation (dotted line).
  Lower plot: Orbital period evolution in rotating and non-rotating system. }
\end{figure}

\begin{figure}
  \centering
   \includegraphics[width=\columnwidth]{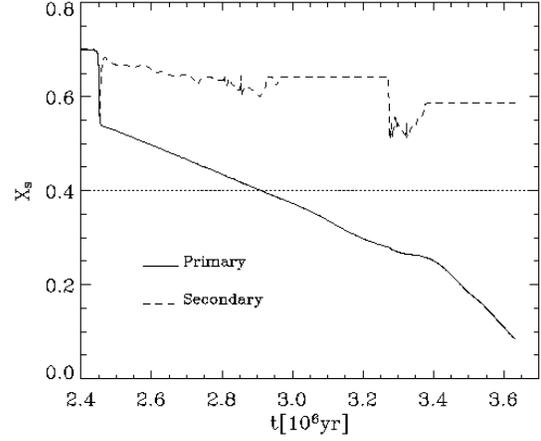}
  \caption{\label{surfrot}The hydrogen surface abundance (solid line) in the primary in system
  with $M_{\rm 1,in}$=56$\,\mathrm{M}_{\odot}$, $M_{\rm 2,in}$=33$\,\mathrm{M}_{\odot}$, $p_{\rm in}$=6 days
  is decreasing during mass transfer
  and further due to stellar wind mass loss. The secondary (dashed
  line) decreases its hydrogen surface abundance due to mass transfer. The 
  dotted line indicates a hydrogen abundance of 0.4, where
  the primary starts losing mass with a WR stellar wind.}
\end{figure}

\begin{figure}
  \centering
   \includegraphics[width=\columnwidth]{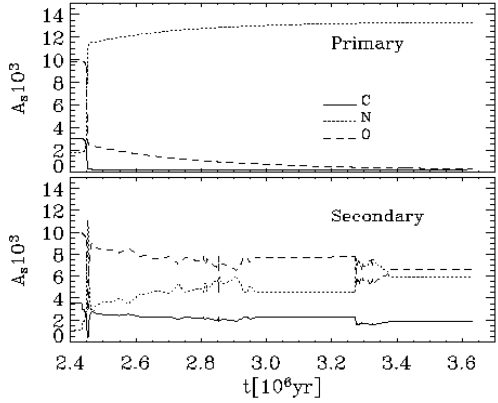}
  \caption{\label{surfcnorot}Surface abundance of carbon (solid line), nitrogen (dotted line) and 
  oxygen (dashed line) in the primary (upper plot) 
  and the secondary (lower plot), in the system with
  $M_{\rm 1,in}$=56$\,\mathrm{M}_{\odot}$, $M_{\rm 2,in}$=33$\,\mathrm{M}_{\odot}$, $p_{\rm in}$=6 days.
  The secondary abundances are changed due to mass transfer of matter from the primary,
  thermohaline mixing and rotational mixing.}
\end{figure}

\begin{figure}
  \centering
   \includegraphics[width=\columnwidth]{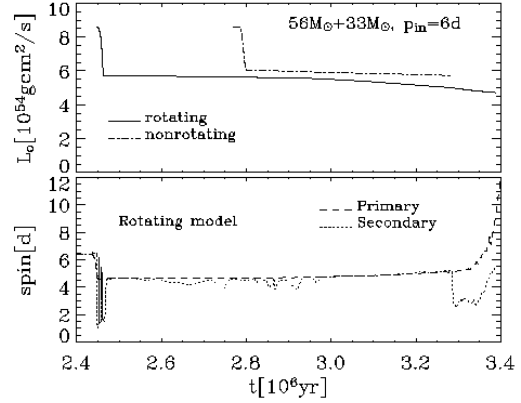}
  \caption{\label{ang}The orbital angular momentum (upper plot) of 
  the non-rotating (dotted line) and the rotating (solid line) binary systems with
  $M_{\rm 1,in}$=56$\,\mathrm{M}_{\odot}$, $M_{\rm 2,in}$=33$\,\mathrm{M}_{\odot}$, $p_{\rm in}$=6 days, decreases
  rapidly due to mass loss from the system during fast Case~A mass transfer and then further due
  to stellar wind mass loss.
  Spin period (lower plot) of the primary (dashed line) and the secondary (dash-dotted line)
  in the above mentioned rotating binary system.}
\end{figure}

\begin{figure}
  \centering
   \includegraphics[width=\columnwidth]{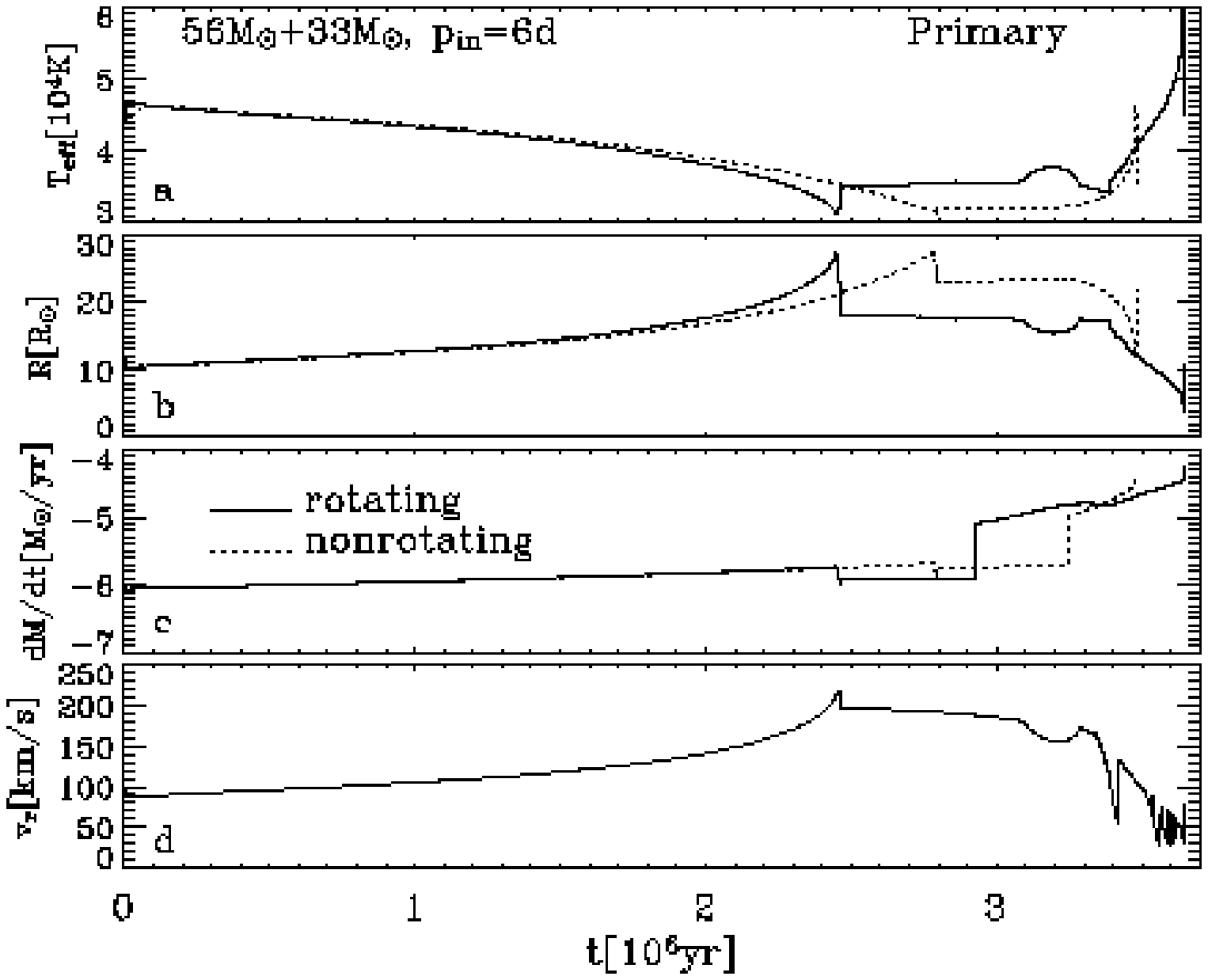}
  \caption{\label{primary}Effective temperature (plot a), stellar radius (plot b), stellar wind mass
  loss rate (plot c) and rotational velocity (plot d) of the primary star 
  in the non-rotating (dotted line) and
  rotating (solid line) binary system with
  $M_{\rm 1,in}$=56$\,\mathrm{M}_{\odot}$, $M_{\rm 2,in}$=33$\,\mathrm{M}_{\odot}$, $p_{\rm in}$=6 days.}
\end{figure}

\begin{figure}
  \centering
   \includegraphics[width=\columnwidth]{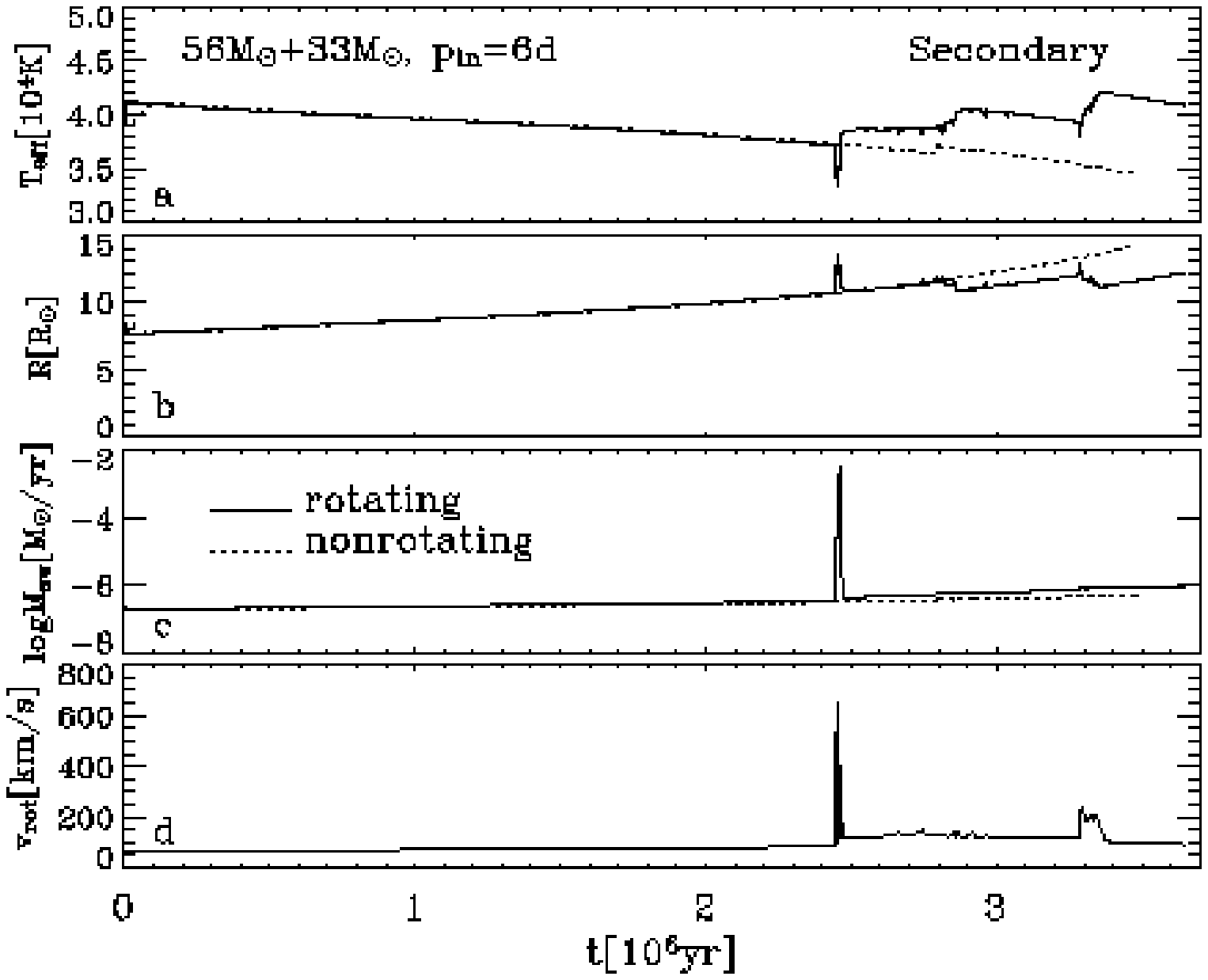}
  \caption{\label{secondary}Effective temperature (plot a), stellar radius (plot b), stellar wind mass
  loss rate (plot c) and rotational velocity (plot d) of the secondary star 
  in the non-rotating (dotted line) and
  rotating (solid line) binary system with
  $M_{\rm 1,in}$=56$\,\mathrm{M}_{\odot}$, $M_{\rm 2,in}$=33$\,\mathrm{M}_{\odot}$, $p_{\rm in}$=6 days.}
\end{figure}

We compare the evolution of non-rotating and 
rotating binary systems on the example $M_{\rm 1,in}$=56$\,\mathrm{M}_{\odot}$ 
$M_{\rm 2,in}$=33$\,\mathrm{M}_{\odot}$, 
an initial orbital period of $p$=6 days, 
and Hamann/3 WR mass loss stellar wind rate (Table~\ref{big2}: N 6).
The rotating binary system is synchronized as it starts core hydrogen burning and it stays that way until
mass transfer starts. The radius of the primary increases during the main sequence phase 
(from $\sim$10 to $\sim$25$\,\mathrm{R}_{\odot}$, Fig.~\ref{primary}b), 
but the rotation of the primary stays synchronized with the orbital
period. This is why the rotational velocity of the primary also increases from $\sim$100 to
$\sim$200$\rm~km~s^{\rm-1}$ (Fig.~\ref{primary}d). The radius of the rotating primary increases faster
than the radius of the non-rotating primary due to the influence of the centrifugal force.  
The result is that Case~A mass transfer starts earlier for the rotating binary system
($X_{\rm c,non}$$\approx$80\%) then for the corresponding nonrotating one($X_{\rm c,rot}$$\approx$71\%,
Fig.~\ref{mtrrot}).

When the fast phase of Case~A starts, the secondary spins up (Fig.~\ref{secondary}d) 
and stellar wind mass loss rapidly increases 
($\dot M_{\rm sw}$$\sim$10$^{\rm -3} \,\mathrm{M}_{\odot}\rm~yr^{\rm -1}$, Fig.~\ref{secondary}c). 
The accretion efficiency during this phase in the rotating system is $\beta$=0.15 (Table~\ref{big2}).
We see in Fig.~\ref{mtrrot} that the orbital period after Case~A mass transfer of the 
rotating binary system is
shorter than for the non-rotating system (4.5 compared with 6.6 days).
The orbital angular momentum of the binary is changing due to mass transfer, 
mass loss from the system and spin-orbit coupling. The rotating binary system loses more angular
momentum and the final orbital period is shorter than in the corresponding non-rotating system.
Angular momentum loss in our systems is calculated according to \citet{1992ApJ...391..246P} 
as already mentioned
in Sect~\ref{code}, and parameter $\alpha$ that determines the efficiency of angular momentum loss is
calculated according to \citet{1993ApJ...410..719B}. It increases with the mass ratio $M_{\rm 2}/M_{\rm 1}$ and 
the ratio between the secondary radius and its Roche radius $R_{\rm 2}/R_{\rm l2}$. In rotating system the secondary
accretes slightly more matter ($\bar \beta$=0.15) compared to $\beta$=0.1 in non-rotating systems,
so the mass ratio $M_{\rm 2}/M_{\rm 1}$ is larger in the rotating system. Second, the secondary is spinning fast and its
radius is larger than in the non-rotating case, and so is the ratio $R_{\rm 2}/R_{\rm l2}$. The result is that the angular
momentum is more efficiently removed from the system in the rotating binary system.
After the fast phase of Case~A mass transfer, the two primaries, non-rotating and rotating, have almost the same
mass $\sim$34$\,\mathrm{M}_{\odot}$ and helium surface abundance $Y_{\rm s}$$\sim$44\%. However, since the orbital
periods are different, so are the radii of the
primaries ($\sim$18$\,\mathrm{R}_{\odot}$ for the rotating and
$\sim$23$\,\mathrm{R}_{\odot}$ for the non-rotating
case).

When the fast phase of Case~A is finished, the non-rotating primary has still $\sim$20\% of
hydrogen to burn ($\sim$7$\cdot$10$^{\rm 5} \rm~yr$), and the rotating primary has 
$\sim$10\% more than
that ($\sim$1.2$\cdot$10$^{\rm 6} \rm~yr$). When the surface hydrogen abundance is less than $40$\%,
the primaries start losing mass as WR stars, i.e., their stellar wind mass loss rate increases.
Since the rotating primary has more time to spend on the main sequence, it also has more time to lose
mass by WR stellar wind mass loss (7.2$\cdot$10$^{\rm 5} \rm~yr$ compared with 2.5$\cdot$10$^{\rm 5} \rm
~yr$ for non-rotating system).The result is that the non-rotating primary enters Case~AB 
mass transfer as a
$\sim$26$\,\mathrm{M}_{\odot}$ star with $Y_{\rm s}$=0.75, while the rotating one is a $\sim$17$\,\mathrm{M}_{\odot}$ star
with $Y_{\rm s}$=0.90. Clearly, the rotating primary has less hydrogen in its envelope, i.e. less mass to
transfer to the secondary during Case~AB mass transfer, and the orbit widens less than in the non-rotating
system.
We can draw the conclusion that if rotation is included in our calculations, 
the initial WR mass is smaller and the orbital period of the WR+O system is
shorter than in the corresponding non-rotating system (Table~\ref{mala}).

We present in Fig.~\ref{hrrot} the evolutionary tracks of the rotating primary and secondary in the 
HR diagram.
Both stars are core hydrogen burning stars (dashed line, Fig.~\ref{hrrot}), 
but since the primary is more
massive, it evolves faster and fills its Roche lobe, so 
the system enters Case~A mass transfer (solid line, Fig.~\ref{hrrot}). 
The primary loses matter quickly with a high mass transfer rate
($\dot M_{\rm tr}^{\rm \rm max}$$\approx$3.2$\cdot$10$^{\rm
-3}\,\mathrm{M}_{\odot}\rm~yr^{\rm -1}$) 
and its luminosity decreases (Fig.~\ref{hrrot}). 
At the same time 
the secondary accretes matter and its luminosity increases, but due to
change in rotational velocity (Fig.~\ref{secondary}d) its radius and effective temperature are changing
as well (Fig.~\ref{hrrot}d, Fig.~\ref{secondary}a,b). 
During fast Case~A mass transfer the primary lost $\sim$19$\,\mathrm{M}_{\odot}$ and the secondary
accreted $15$\% of that matter. 
After the fast mass transfer, 
the primary is still burning hydrogen in
its core and is still expanding, so  
slow Case~A mass transfer takes place. After the primary starts losing mass with a WR
stellar wind mass loss rate ($X_{\rm s}$$<$0.4) its radius will decrease and the slow phase of 
Case~A stops (Fig.~\ref{primary}c). 
However, the primary continues expanding on the nuclear time scale (Fig.~\ref{primary}b)
and it fills its Roche lobe once again (Fig.~\ref{mtrrot}, upper plot). 
At the end of core hydrogen burning the primary contracts (effective temperature increases) 
and thus RLOF stops. This phase is presented in Fig.~\ref{hrrot} with a dotted line. 
When hydrogen starts burning in
a shell, the primary star expands (dash-dotted line, Fig.~\ref{hrrot}), 
fills its Roche lobe and Case~AB mass transfer starts.

The initial helium core masses are 18.6$\,\mathrm{M}_{\odot}$ for the non-rotating and 14.8$\,\mathrm{M}_{\odot}$ for
the rotating primary. When Case~AB mass transfer starts, the orbital periods are 7.9 d and 6.6 d
for the non-rotating and the rotating system respectively (Fig.~\ref{mtrrot}, lower plot).
The non-rotating primary loses $\sim$7$\,\mathrm{M}_{\odot}$ and the rotating one $\sim$2$\,\mathrm{M}_{\odot}$
during Case~AB. When there is more mass to be transfered from the less to the more massive star in a binary
system, the orbit widens more and the final orbital period is longer. 

Fig.~\ref{conv1rot} and Fig.~\ref{conv2rot} show the structure of the primary and the secondary 
before Case~AB mass transfer.
The primary loses large amounts of matter during the fast phase of Case~A mass transfer ($\sim$20$\,\mathrm{M}_{\odot}$), 
and its convective core becomes less than half of its original mass. At the same time, the
secondary accretes matter from the primary and the heavier elements are being relocated by thermohaline
mixing. 
Fig.~\ref{surfrot} and Fig.~\ref{surfcnorot} 
show surface abundances of the primary and the secondary.
The secondary is accreting material from the primary and its surface abundances 
change due to this, but also due to thermohaline and rotational mixing.

Fig.~\ref{ang} shows the orbital angular momentum of the system and the spin periods 
of both components. The orbital angular momentum of the system
decreases rapidly due to mass loss from the system during fast Case~A mass transfer, and then further due
to stellar wind mass loss.
The primary slows down rapidly during fast Case~A and
further due to stellar wind mass loss. The secondary spins up due to the accretion from the primary
during fast Case~A mass transfer and then slows down due to stellar wind mass loss. It
spins up again during slow Case~A mass transfer.

The masses of modelled WR stars are in the range from $\sim$11$\,\mathrm{M}_{\odot}$ 
to $\sim$15.7$\,\mathrm{M}_{\odot}$. Period of modelled WR+O systems vary 
from $\sim$7.6 to
$\sim$12.7 days and mass ratios are between 0.35 and 0.46 (Table~\ref{big2}).

\begin{table}
\caption[]{\label{mala}Comparison of resulting WR masses and orbital periods from 
non-rotating and rotating binary systems with the same initial parameters.$M_{\rm 1,in},M_{\rm 2,in}$ are
initial primary and secondary mass, $p_{\rm in}$ is the initial orbital period,
$M_{\rm WR,5}$, $M_{\rm WR,1}$ are WR masses at $X_{\rm s}=0.05$ and $X_{\rm s}\le0.01$ respectively
and $p$ is the orbital period in the initial WR+O system where the hydrogen surface abundance of WR star is 
$X_{\rm s}=0.05$. Systems are modelled with WR stellar wind mass loss H/6 except $^{\rm *}$ 
which are done with H/3, $^R$ indicates rotating models.}
\centerline{}
\begin{tabular*}{\linewidth}{@{\extracolsep{\fill}}lcccc}
\hline
\hline
\\
$M_{\rm 1,in}+M_{\rm 2,in}$ & $p_{\rm in}$ & $M_{\rm WR,5}$ & $M_{\rm WR,1}$ & $p$\\
$[~\,\mathrm{M}_{\odot}]$ &  $[\rm d]$ & $[~\,\mathrm{M}_{\odot}]$ & $[~\,\mathrm{M}_{\odot}]$ & $\rm [d]$ \\   
\\      
\hline
$41+20$ &    $6$ &  $11.8$  &    $11.2$  & $12.6$\\     
$41+20^{\rm R}$ &  $6$ &  $11.0$  &    $10.2$ & $9.8$\\     
$41+24$ &    $6$ &  $12.1$  &    $11.4$  & $21.5$\\     
$41+24^{\rm R}$ &  $6$ &  $11.2$  &    $10.5$ & $16.4$\\     
$56+33^{\rm *}$ &    $6$ &  $18.6$  &    $17.5$ & $13.8$\\ 
$56+33^{\rm *,R}$ &  $6$ &  $14.9$  &    $13.6$ & $8.5$\\    
\hline
\hline
\end{tabular*}
\end{table}

\subsection{Influence of rotation on the accretion efficiency}\label{acc}

We show in Table~\ref{beta} average accretion efficiencies of rotating binary systems during different
mass transfer phases, and total average values with and without stellar wind mass loss from the primary
included.
During fast Case~A mass transfer, the primary stars are losing matter with very high mass transfer rates 
(3..6.5$\cdot$10$^{-3} \,\mathrm{M}_{\odot}\rm~yr^{\rm -1}$). The angular momentum of surface layers in the secondary 
increases fast, they spin up to close to the critical rotation and start losing mass 
with high mass loss rate 
($\sim$10$^{-3} \,\mathrm{M}_{\odot}$). The average accretion efficiency during fast Case~A in our
models is 15-20\%. Since this phase takes place on the thermal time scale, stellar wind mass loss from the
primary is negligible during this phase.

Slow Case~A mass transfer takes place on the nuclear time scale. The primary stars start losing
their mass due to a WR stellar wind when their surfaces become hydrogen deficient ($X_{\rm s}$$<$0.4). 
The WR stellar wind mass
loss rates are of the order of : $\dot M\sim$10$^{\rm -5}
\,\mathrm{M}_{\odot}\rm~yr^{\rm -1}$, and
we have to take into account stellar wind mass loss of the primary during slow Case~A.
We calculate the mass loss of the primary only due to mass transfer and total mass loss including stellar
wind mass loss, and the two corresponding average accretion efficiencies.
If we calculate $\bar \beta_{\rm s}$ only for mass transfer, we notice that the slow Case~A is 
almost
a conservative process. The average mass transfer rates are $\sim$10$^{\rm -6}
\,\mathrm{M}_{\odot}\rm~yr^{\rm -1}$
and the secondary stars are able to accrete almost everything without spinning up to critical
rotation.

\begin{figure}
  \centering
   \includegraphics[width=\columnwidth]{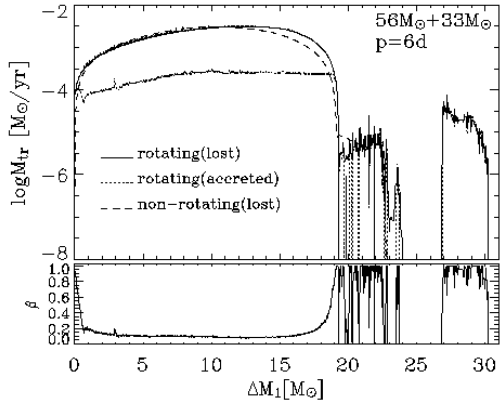}
  \caption{\label{mbeta}Upper plot: Mass transfer (solid line) and accretion rate (dotted line) of the rotating 
  initial system 
  56$\,\mathrm{M}_{\odot}$+33$\,\mathrm{M}_{\odot}$, $p$=6 days. Dashed line represent mass transfer rate in the corresponding
  non-rotating binary.
  Lower plot: accretion efficiency of the secondary taking into 
  account matter lost by the primary only due to mass transfer.}
  \label{fig6}
\end{figure}

Fig.~\ref{mbeta} shows how mass transfer rate, accretion rate and  $\beta$ 
change in the rotating model with 56$\,\mathrm{M}_{\odot}$+33$\,\mathrm{M}_{\odot}$, $p_{\rm in}$=6 days 
(WR mass loss Hamann/3) depending on the amount of matter lost by the primary. 
We also see in this figure the mass transfer rate from the primary in the non-rotating case.
We can notice in the upper plot, what we previously discussed,
that during most of the fast Case~A mass transfer, 
the mass accretion rate of the secondary is about
one order of magnitude lower than the mass loss rate of the primary.
The primary loses $\sim$19.3$\,\mathrm{M}_{\odot}$ during the fast phase and the secondary gains
$\sim$2.9$\,\mathrm{M}_{\odot}$, which means that on
average $\sim$15\% of the mass has been accreted. 
However, the mass loss of the primary due to mass transfer during the slow phase is 
$\sim$4.5$\,\mathrm{M}_{\odot}$, and the secondary accretes $\sim$3.9$\,\mathrm{M}_{\odot}$ which means that
$\bar \beta_{\rm s}\sim$0.87.
If we take into account stellar wind mass loss of the primary stars, 
the average accretion efficiencies are lower. 
For example, the total mass loss of the
primary during slow Case~A mass transfer, including the stellar wind, in the previous example is 
$\sim$10.9$\,\mathrm{M}_{\odot}$, which means that ${\bar \beta_{\rm s}}^{\rm w}$$\approx$0.36.
We neglected accretion during Case~AB mass transfer since \citet{wellsteinphd} showed that
it is inefficient, and since the primary stars in the modelled systems 
have relatively low mass hydrogen envelopes, and masses of secondary stars will not 
significantly change due to this mass transfer.
(However, let us not forget that even the accretion of very small amounts of matter 
can be important for spinning up the secondary's surface layers and making it rotate
faster than synchronously in a WR+O binary system.)
Also, since this mass transfer
takes place on the thermal time scale, stellar wind mass loss can be neglected.

Finally, we can estimate the total mass loss from the binary systems including stellar wind, 
or only due to mass
transfer, and calculate corresponding values of $\beta$.
In the binary systems we modelled, the primary stars lose between 30$\,\mathrm{M}_{\odot}$ 
and 45$\,\mathrm{M}_{\odot}$ due to mass
transfer and stellar wind, until they
ignite helium in their core. The amount of lost mass increases with initial mass. At the same time
the secondaries accrete 3..10$\,\mathrm{M}_{\odot}$. 
This means that in most cases 80..90\% of the mass lost by the primary leaves the binary system. 
On the other hand, the primary stars lose $\sim$20..30$\,\mathrm{M}_{\odot}$ only due to mass transfer, so
the average accretion of secondary stars in our models is between 15 and 30\%.

\begin{table}
\caption[]{\label{beta}Mass loss from binary systems. $N$ is number of the model corresponding to
Table~\ref{big2}. $\bar \beta_{\rm fast}$ is the average accretion efficiency of the secondary during the 
fast phase of
Case~A mass transfer. $\bar \beta_{\rm slow}$ is the accretion efficiency of the secondary during the 
slow phase of
Case~A mass transfer taking into account matter lost by the primary only due to the mass transfer. 
${\bar \beta_{\rm slow}}^{\rm wind}$
is the average accretion efficiency of the secondary during the slow phase of
Case~A mass transfer taking into account matter lost by the primary due to the mass transfer and stellar
wind. $\bar \beta$ is the average accretion efficiency of the secondary during the 
progenitor evolution of WR+O binary
system taking into account matter lost by the primary only due to the mass transfer and 
${\bar \beta_{\rm mtr}}^{\rm wind}$ taking also into account stellar wind mass loss of the primary.}
\centerline{}
\begin{tabular*}{\linewidth}{@{\extracolsep{\fill}}lccccc}
\hline
\hline
\\
$Nr$ & $\bar \beta_{\rm fast}$ & $\bar \beta_{\rm slow}$ & ${\bar \beta_{\rm slow}}^{\rm wind}$ & 
$\bar \beta$ & $\bar \beta^{wind}$
\\
\hline
\\
$R1  $ & $0.18$ & $0.80$ & $0.35$ & $0.15$ & $0.13$\\  
$R2  $ & $0.18$ & $0.95$ & $0.65$ & $0.15$ & $0.10$\\  
$R3  $ & $0.33$ & $0.90$ & $0.80$ & $0.40$ & $0.37$\\  
$R4  $ & $0.23$ & $0.94$ & $0.60$ & $0.20$ & $0.18$\\  
$R5  $ & $0.15$ & $0.90$ & $0.58$ & $0.28$ & $0.22$\\  
$R6  $ & $0.15$ & $0.87$ & $0.36$ & $0.23$ & $0.14$\\  
$R7  $ & $0.15$ & $0.88$ & $0.70$ & $0.18$ & $0.09$\\  
$R8  $ & $0.20$ & $0.88$ & $0.58$ & $0.30$ & $0.23$\\  
$R9  $ & $0.20$ & $0.89$ & $0.40$ & $0.30$ & $0.16$\\  
\hline
\hline
\end{tabular*}
\end{table}

%%%%%%%%%%%%%%%%%%%%%%%%%%%%%%%%%%%%%%%%%%%%%%%%%%%%%%%%%%%%%%%%%%%%%%%%%%%%%%
\section{Comparison with observations}\label{comp}

Our rotating models give generally similar results as our non-rotating models 
for $\beta=0.1$. 

The rotating binary systems  R6 (56$\,\mathrm{M}_{\odot}$+33$\,\mathrm{M}_{\odot}$, $p$=6 days,  
Hamann/3 WR mass loss) and R1 and R2
(41$\,\mathrm{M}_{\odot}$+20$\,\mathrm{M}_{\odot}$, $p$=6 days) agree quite well with the observed 
systems HD\,186943 and HD\,90657,
as well as the non-rotating systems N11 and N12
(56$\,\mathrm{M}_{\odot}$+33$\,\mathrm{M}_{\odot}$, $p$=6 days; WR mass loss rate Hamann/2 and
Hamann/3). 
The system R6 
 evolves into a WR+O configuration with
15$\,\mathrm{M}_{\odot}$+39$\,\mathrm{M}_{\odot}$ and $p$=8.5 days. 
I.e., its masses and period are close to those found in HD\,186943 and HD\,90657,
even though its mass ratio of 0.38 is somewhat smaller than what is observed.
Systems R1 and R2
evolve into a
11$\,\mathrm{M}_{\odot}$+24$\,\mathrm{M}_{\odot}$ WR+O system with a 9.8~day orbital period.
I.e., period and mass ratio (0.46) agree well with the observed systems, 
but the stellar masses are somewhat smaller than observed (cf. Sect.~\ref{obs}).
Systems N11 and N12 evolve into a WR+O system of
19$\,\mathrm{M}_{\odot}$+35$\,\mathrm{M}_{\odot}$ with an orbital period of 12..14 days.
In this case, both masses and the mass ratio (0.53) agree well with the observed
ones, but the orbital period is slightly too large.
I.e., although none of our models is a perfect match of HD\,186943 or HD\,90657
--- which to find would require many more models, however, might not teach us
very much --- it is clear form these results that both systems can in fact be well
explained through highly inefficient Case~A mass transfer.

The situation is more difficult with HD\,211853 (GP Cep):
neither the models with nor those without rotation reproduce it satisfactory.
HD\,211853 has the shortest period (6.7~d) and largest mass ratio (0.54) of the three
chosen Galactic WR+O binaries. While we can not exclude that
a Case~A model of the kind presented here can reproduce this systems, 
especially the small period makes it appear more likely that this
system has gone through a contact phase: contact would reduce the orbital angular momentum,
and increase the mass loss from the system, i.e. result in a larger WR/O mass ratio
\citep{2001A&A...369..939W}. 
This reasoning is strengthened by the consideration that, in contrast to HD\,186943 or HD\,90657,
the WR star in HD\,211853 is of spectral type WN6/WNC. I.e., as this spectroscopic
signature is not interpreted in terms of a binary nature of the WR component,
but rather by assuming that the WR star is in the transition phase from the WN to the
WC stage \citep{1989ApJ...344..870M,1991A&A...252..669L}. This implies that the WR star 
in HD\,211853 must have already lost several solar masses of helium-rich matter,
which causes the orbit to widen. 
For example, system R6, which evolved into a 
14.8$\,\mathrm{M}_{\odot}$+39.0$\,\mathrm{M}_{\odot}$ WR+O system with $p$=8.53 days, 
evolves into a WC+O system 
after losing $\sim$5$\,\mathrm{M}_{\odot}$ more from the Wolf-Rayet star,
which increases its orbital period by $\sim$3 days.
I.e., HD\,211853 might have entered the WR+O stage with an orbital period of about 
4~days, which would put it together with the shortest period WR binaries like
CX Cep or CQ Cep whose periods are 2.1 and 1.64 days respectively.

During the evolution of WR+O binary system, the primary loses
mass due to WR stellar wind mass loss.
WR stellar wind mass loss of the primary
decreases mass ratio of the system and increases the orbital period, which means
that, for example, WC+O binary system HD63099 ($M_{\rm WR}$=9$\,\mathrm{M}_{\odot}$,
$M_{\rm O}$=32$\,\mathrm{M}_{\odot}$ and $p$=14 days) could have evolved into present state
through a WN+O binary system with $q$=0.5.

%%%%%%%%%%%%%%%%%%%%%%%%%%%%%%%%%%%%%%%%%%%%%%%%%%%%%%%%%%%%%%%%%%%%%%%%%%%%%%%
%\section{Further evolution of WR+O system}\label{further}
%WR+O phase in our models starts after Case~AB mass transfer, when the primary has less than $5$\% of
%hydrogen on the surface and it is burning helium in the core. At the same time, the secondary is still 
%main sequence O star. During WR+O phase, the primary is burning helium in the core 
%and losing mass with high mass loss
%rates. Due to this mass loss from the system, binary orbit widens.

%After this phase,
%primary goes through helium shell burning phase, followed by carbon and oxygen core burning.
%When the primary explodes in SN, it has mass $\sim4.5\,\mathrm{M}_{\odot}$ star, the secondary is main sequence
%star $38.6 \,\mathrm{M}_{\odot}$ and orbital period is $p=13.5$ days.
%If the binary system is not disrupted by SN explosion, it will become 
%neutron star+main sequence star binary. In further evolution, the secondary will become WR star, then
%also explode as a SN and if the system is not disrupted after second explosion, it will become double
%neutron star binary.

%%%%%%%%%%%%%%%%%%%%%%%%%%%%%%%%%%%%%%%%%%%%%%%%%%%%%%%%%%%%%%%%%%%%%%%%%%%%%%
\section{Conclusions}\label{concl}

%We calculated evolution of nonconservative nonrotating 
%($\beta=0.1$) Case~A binary systems with
%primaries $M_{\rm 1,i}=41..56 \,\mathrm{M}_{\odot}$ until WR+O stage. 
%We estimated the relation between the
%initial primary mass and the initial WR mass for systems with the initial period $3$ days
%and the initial mass ratio $q\approx1.7$.
%We conclude that the initial WR mass increases if the initial mass ratio ($M_{\rm 1,in}/M_{\rm 2,in})$
%or stellar wind mass loss decreases or initial orbital period increases. 
%We also find out that orbital period in WR+O system increases if the initial mass
%ratio or mass loss rate decreases.

In an effort to constrain the progenitor evolution of the three WN+O
binaries HD\,186943, HD\,90657, and HD\,211853,
we calculated the evolution of non-conservative Case~A binary systems with
primaries $M_{\rm 1,i}$=41..65$\,\mathrm{M}_{\odot}$ and initial mass ratios
between 1.7 and 2 until the WN+O stage.
We performed binary evolution calculations neglecting rotational processes in the
two stellar components, and assuming a constant mass accretion efficiency of
10\% for all three phases of the mass transfer, fast Case~A, slow Case~A, and Case~AB.
Those models could match two of the three systems reasonably well,
while HD\,211853, which has the shortest orbital period, the largest mass ratio, and
a WN/WC Wolf-Rayet component, was found to be not well explained by contact-free 
evolutionary models: While models with shorter initial orbital periods result in short periods 
during the WR+O stage, the initial WR mass is decreasing at the same time, which
leads to smaller initial WR/O mass ratios.

We then computed binary evolution models including the physics of rotation in
both stellar components as well as the spin-up process of the mass gainer due
to angular momentum accretion. In these models, the surface of the accreting star
is continuously spun-up by accretion, while at the same time angular momentum
is transported from the outer layers into the stellar interior by rotationally
induced mixing processes. By employing a simple model for the mass loss of
rapidly rotating luminous stars -- the so called $\Omega$-limit, which 
was actually worked out to describe the mass loss processes in Luminous Blue Variables
\citep{1997lbv..conf...83L} --- accretion is drastically reduced once the star reaches 
critical rotation at its surface. The mass accretion rate is then controlled
by the time scale of internal angular momentum transport.

Some first such model for Case~A and early Case~B have been computed 
by Wellstein \citep{2003IAUS..212..275L,2004IAUS..215c} for a primary mass of 15$\,\mathrm{M}_{\odot}$ and 
a mass ratio close to one. The result was that rather high mass accretion efficiencies
($\beta\simeq$0.7) could be obtained for initial periods shorter than about 8~days.
Here we find that, with the same physical assumptions although at higher system mass,
the accretion efficiency drops to about 10\% at an initial mass ratio of 1.7.
As \citet{wellsteinphd} computed one early Case~A model for a 26$\,\mathrm{M}_{\odot}$+25$\,\mathrm{M}_{\odot}$
system which gave $\beta$=0.63, it is like the high initial mass ratio 
in our models which is responsible for the low accretion efficiency:
Larger initial mass ratios lead to larger mass transfer rates and, as the time scale
of internal angular momentum transport in the accreting star is rather unaffected,
to smaller accretion efficiencies.

Our rotating models --- in which the accretion efficiency is no free
parameter any more but is computed selfconsistent and time-dependent ---
reproduce the observed WR+O binaries quite well, i.e. as good as our
models without rotation physics, where the accretion efficiency is a free
parameter. Our simplified considerations in Sect.~\ref{simple} have shown that this
is unlikely attributable to the freedom in the choice of the initial parameter
of the binary system, i.e. initial masses and period --- at least under the
assumption that contact was avoided. In case of contact, various new parameters
enter the model, similar to the case of common envelope evolution.
And indeed, also our rotating models can not reproduce  HD\,211853
very well, mostly because it currently has a too short orbital period, which was
likely even significantly shorter at the beginning of its WR+O stage.
However, this of course only confirms the result of the simpler approaches
that a contact-free approach does not work well for this system.

In summary we can say that the system mass and angular momentum loss model used here
--- which is the first detailed approach to tackle the long-standing angular
momentum problem in mass transferring binaries --- has passed the test
of WR+O binaries. However, it still needs to be explored over which
part of the space spanned by the initial binary parameters this model 
works well, and to what extent its results are sensitive to future 
improvements in the stellar interior physics. The inclusion of magnetic
fields generated by differential rotation \citep{2002A&A...381..923S} will be the next
step in this direction \citep{gammapaper}.

%%%%%%%%%%%%%%%%%%%%%%%%%%%%%%%%%%%%%%%%%%%%%%%%%%%%%%%%%%%%%%%%%%%%%%%%%%%
\bibliographystyle{aa}
\bibliography{2368wro}

\begin{thebibliography}{55}
\expandafter\ifx\csname natexlab\endcsname\relax\def\natexlab#1{#1}\fi

\bibitem[{{Bonanos} {et~al.}(2004){Bonanos}, {Stanek}, {Udalski},
  {Wyrzykowski}, {{\. Z}ebru{\' n}}, {Kubiak}, {Szyma{\' n}ski}, {Szewczyk},
  {Pietrzy{\' n}ski}, \& {Soszy{\' n}ski}}]{2004ApJ...611L..33B}
{Bonanos}, A.~Z., {Stanek}, K.~Z., {Udalski}, A., {et~al.} 2004, \apjl, 611,
  L33

\bibitem[{{Braun}(1998)}]{1998PhDT........29B}
{Braun}, A. 1998, Ph.D.~Thesis

\bibitem[{{Braun} \& {Langer}(1995)}]{1995A&A...297..483B}
{Braun}, H. \& {Langer}, N. 1995, \aap, 297, 483

\bibitem[{{Brookshaw} \& {Tavani}(1993)}]{1993ApJ...410..719B}
{Brookshaw}, L. \& {Tavani}, M. 1993, \apj, 410, 719

\bibitem[{{Chevalier} \& {Ilovaisky}(1998)}]{1998A&A...330..201C}
{Chevalier}, C. \& {Ilovaisky}, S.~A. 1998, \aap, 330, 201

\bibitem[{{de Loore} \& {de Greve}(1992)}]{1992A&AS...94..453D}
{de Loore}, C. \& {de Greve}, J.~P. 1992, \aaps, 94, 453

\bibitem[{{Demers} {et~al.}(2002){Demers}, {Moffat}, {Marchenko}, {Gayley}, \&
  {Morel}}]{2002ApJ...577..409D}
{Demers}, H., {Moffat}, A.~F.~J., {Marchenko}, S.~V., {Gayley}, K.~G., \&
  {Morel}, T. 2002, \apj, 577, 409

\bibitem[{{Dessart} {et~al.}(2003){Dessart}, {Langer}, \&
  {Petrovic}}]{2003A&A...404..991D}
{Dessart}, L., {Langer}, N., \& {Petrovic}, J. 2003, \aap, 404, 991

\bibitem[{{Hamann} \& {Koesterke}(1998)}]{1998A&A...335.1003H}
{Hamann}, W.-R. \& {Koesterke}, L. 1998, \aap, 335, 1003

\bibitem[{{Hamann} {et~al.}(1995){Hamann}, {Koesterke}, \&
  {Wessolowski}}]{1995A&A...299..151H}
{Hamann}, W.-R., {Koesterke}, L., \& {Wessolowski}, U. 1995, \aap, 299, 151

\bibitem[{{Heger} \& {Langer}(2000)}]{2000ApJ...544.1016H}
{Heger}, A. \& {Langer}, N. 2000, \apj, 544, 1016

\bibitem[{{Heger} {et~al.}(2000){Heger}, {Langer}, \&
  {Woosley}}]{2000ApJ...528..368H}
{Heger}, A., {Langer}, N., \& {Woosley}, S.~E. 2000, \apj, 528, 368

\bibitem[{{Heger} {et~al.}(2004){Heger}, {Woosley}, {Langer}, \&
  {Spruit}}]{2004IAUS..215b}
{Heger}, A., {Woosley}, S.~E., {Langer}, N., \& {Spruit}, H.~C. 2004, in IAU
  Symposium 215, 591

\bibitem[{{Iglesias} \& {Rogers}(1996)}]{1996ApJ...464..943I}
{Iglesias}, C.~A. \& {Rogers}, F.~J. 1996, \apj, 464, 943

\bibitem[{{King} {et~al.}(2001){King}, {Schenker}, {Kolb}, \&
  {Davies}}]{2001MNRAS.321..327K}
{King}, A.~R., {Schenker}, K., {Kolb}, U., \& {Davies}, M.~B. 2001, \mnras,
  321, 327

\bibitem[{{Kippenhahn} {et~al.}(1967){Kippenhahn}, {Kohl}, \&
  {Weigert}}]{1967ZA.....66...58K}
{Kippenhahn}, R., {Kohl}, K., \& {Weigert}, A. 1967, Zeitschrift fur
  Astrophysics, 66, 58

\bibitem[{{Kippenhahn} \& {Thomas}(1970)}]{1970stro.coll...20K}
{Kippenhahn}, R. \& {Thomas}, H.-C. 1970, in IAU Colloq. 4: Stellar Rotation,
  20

\bibitem[{{Kopal}(1978)}]{1978dcbs.conf.....K}
{Kopal}, Z., ed. 1978, {Dynamics of Close Binary Systems}

\bibitem[{{Kudritzki} {et~al.}(1989){Kudritzki}, {Pauldrach}, {Puls}, \&
  {Abbott}}]{1989A&A...219..205K}
{Kudritzki}, R.~P., {Pauldrach}, A., {Puls}, J., \& {Abbott}, D.~C. 1989, \aap,
  219, 205

\bibitem[{{Lamontagne} {et~al.}(1996){Lamontagne}, {Moffat}, {Drissen},
  {Robert}, \& {Matthews}}]{1996AJ....112.2227L}
{Lamontagne}, R., {Moffat}, A.~F.~J., {Drissen}, L., {Robert}, C., \&
  {Matthews}, J.~M. 1996, \aj, 112, 2227

\bibitem[{{Langer}(1991)}]{1991A&A...252..669L}
{Langer}, N. 1991, \aap, 252, 669

\bibitem[{{Langer}(1997)}]{1997lbv..conf...83L}
{Langer}, N. 1997, in ASP Conf. Ser. 120: Luminous Blue Variables: Massive
  Stars in Transition, 83

\bibitem[{{Langer}(1998)}]{1998A&A...329..551L}
---. 1998, \aap, 329, 551

\bibitem[{{Langer} {et~al.}(2000){Langer}, {Deutschmann}, {Wellstein}, \& {H{\"
  o}flich}}]{2000A&A...362.1046L}
{Langer}, N., {Deutschmann}, A., {Wellstein}, S., \& {H{\" o}flich}, P. 2000,
  \aap, 362, 1046

\bibitem[{{Langer} {et~al.}(2003){Langer}, {Wellstein}, \&
  {Petrovic}}]{2003IAUS..212..275L}
{Langer}, N., {Wellstein}, S., \& {Petrovic}, J. 2003, in IAU Symposium 212,
  275

\bibitem[{{Langer} {et~al.}(2004){Langer}, {Yoon}, {Petrovic}, \&
  {Heger}}]{2004IAUS..215c}
{Langer}, N., {Yoon}, S.-C., {Petrovic}, J., \& {Heger}, A. 2004, in IAU
  Symposium 215, 535

\bibitem[{{Maeder} \& {Meynet}(2003)}]{2003A&A...411..543M}
{Maeder}, A. \& {Meynet}, G. 2003, \aap, 411, 543

\bibitem[{{Massey}(1981)}]{1981ApJ...244..157M}
{Massey}, P. 1981, \apj, 244, 157

\bibitem[{{Massey} \& {Grove}(1989)}]{1989ApJ...344..870M}
{Massey}, P. \& {Grove}, K. 1989, \apj, 344, 870

\bibitem[{{Meyer} \& {Meyer-Hofmeister}(1983)}]{1983A&A...121...29M}
{Meyer}, F. \& {Meyer-Hofmeister}, E. 1983, \aap, 121, 29

\bibitem[{{Meynet} \& {Maeder}(2000)}]{2000A&A...361..101M}
{Meynet}, G. \& {Maeder}, A. 2000, \aap, 361, 101

\bibitem[{{Niemela} \& {Moffat}(1982)}]{1982ApJ...259..213N}
{Niemela}, V.~S. \& {Moffat}, A.~F.~J. 1982, \apj, 259, 213

\bibitem[{{Packet}(1981)}]{1981A&A...102...17P}
{Packet}, W. 1981, \aap, 102, 17

\bibitem[{{Paczy{\' n}ski}(1967)}]{1967AcA....17..355P}
{Paczy{\' n}ski}, B. 1967, Acta Astronomica, 17, 355

\bibitem[{{Paczynski}(1991)}]{1991ApJ...370..597P}
{Paczynski}, B. 1991, \apj, 370, 597

\bibitem[{{Panov} \& {Seggewiss}(1990)}]{1990A&A...227..117P}
{Panov}, K.~P. \& {Seggewiss}, W. 1990, \aap, 227, 117

\bibitem[{{Petrovic} \& {Langer}(2004)}]{2004RMxAC..20..231P}
{Petrovic}, J. \& {Langer}, N. 2004, in Revista Mexicana de Astronomia y
  Astrofisica Conference Series, 231--231

\bibitem[{{Petrovic} {et~al.}(2004){Petrovic}, {Langer}, {Yoon}, \&
  {Heger}}]{gammapaper}
{Petrovic}, J., {Langer}, N., {Yoon}, S.-C., \& {Heger}, A. 2004, \aap,
  accepted

\bibitem[{{Podsiadlowski} {et~al.}(1992){Podsiadlowski}, {Joss}, \&
  {Hsu}}]{1992ApJ...391..246P}
{Podsiadlowski}, P., {Joss}, P.~C., \& {Hsu}, J.~J.~L. 1992, \apj, 391, 246

\bibitem[{{Rauw} {et~al.}(2004){Rauw}, {De Becker}, {Naz{\' e}}, {Crowther},
  {Gosset}, {Sana}, {van der Hucht}, {Vreux}, \&
  {Williams}}]{2004A&A...420L...9R}
{Rauw}, G., {De Becker}, M., {Naz{\' e}}, Y., {et~al.} 2004, \aap, 420, L9

\bibitem[{{Ritter}(1988)}]{1988A&A...202...93R}
{Ritter}, H. 1988, \aap, 202, 93

\bibitem[{{Spruit}(2002)}]{2002A&A...381..923S}
{Spruit}, H.~C. 2002, \aap, 381, 923

\bibitem[{{Underhill} {et~al.}(1988){Underhill}, {Yang}, \&
  {Hill}}]{1988PASP..100.1256U}
{Underhill}, A.~B., {Yang}, S., \& {Hill}, G.~M. 1988, \pasp, 100, 1256

\bibitem[{{van den Heuvel} \& {Heise}(1972)}]{1972NPhS..239...67V}
{van den Heuvel}, E.~P.~J. \& {Heise}, J. 1972, Nature Physical Science, 239,
  67

\bibitem[{{van der Hucht}(2001)}]{2001NewAR..45..135V}
{van der Hucht}, K.~A. 2001, New Astronomy Reviews, 45, 135

\bibitem[{{Vanbeveren}(1982)}]{1982A&A...105..260V}
{Vanbeveren}, D. 1982, \aap, 105, 260

\bibitem[{{Vanbeveren}(1991)}]{1991A&A...252..159V}
---. 1991, \aap, 252, 159

\bibitem[{{Vanbeveren} {et~al.}(1979){Vanbeveren}, {de Greve}, {de Loore}, \&
  {van Dessel}}]{1979A&A....73...19V}
{Vanbeveren}, D., {de Greve}, J.~P., {de Loore}, C., \& {van Dessel}, E.~L.
  1979, \aap, 73, 19

\bibitem[{{Wellstein}(2001)}]{wellsteinphd}
{Wellstein}, S. 2001, Ph.D.~Thesis

\bibitem[{{Wellstein} \& {Langer}(1999)}]{1999A&A...350..148W}
{Wellstein}, S. \& {Langer}, N. 1999, \aap, 350, 148

\bibitem[{{Wellstein} {et~al.}(2001){Wellstein}, {Langer}, \&
  {Braun}}]{2001A&A...369..939W}
{Wellstein}, S., {Langer}, N., \& {Braun}, H. 2001, \aap, 369, 939

\bibitem[{{Woosley}(2004)}]{2004IAUS..215d}
{Woosley}, S. E.~{Heger}, A. 2004, in IAU Symposium 215, 601

\bibitem[{{Yoon} \& {Langer}(2004{\natexlab{a}})}]{2004A&A...419..645Y}
{Yoon}, S.-C. \& {Langer}, N. 2004{\natexlab{a}}, \aap, 419, 645

\bibitem[{{Yoon} \& {Langer}(2004{\natexlab{b}})}]{2004A&A...419..623Y}
---. 2004{\natexlab{b}}, \aap, 419, 623

\bibitem[{{Zahn}(1977)}]{1977A&A....57..383Z}
{Zahn}, J.-P. 1977, \aap, 57, 383

\end{thebibliography}
 
%% %%%%%%%%%%%%%%%%%%%%%%%%%%%%%%%%%%%%%%%%%%%%%%%%%%%%%%%%%%%% End Document
\end{document}